%% file: main.tex
\documentclass[sigplan,screen]{acmart}
\setcopyright{rightsretained}
\acmPrice{}
\acmDOI{10.1145/3519939.3523712}
\acmYear{2022}
\copyrightyear{2022}
\acmSubmissionID{pldi22main-p379-p}
\acmISBN{978-1-4503-9265-5/22/06}
\acmConference[PLDI '22]{Proceedings of the 43rd ACM SIGPLAN International Conference on Programming Language Design and Implementation}{June 13--17, 2022}{San Diego, CA, USA}
\acmBooktitle{Proceedings of the 43rd ACM SIGPLAN International Conference on Programming Language Design and Implementation (PLDI '22), June 13--17, 2022, San Diego, CA, USA}

\usepackage{booktabs}
\usepackage{stackengine}
\usepackage{bussproofs}
\usepackage{colortbl}
\usepackage{caption}
\usepackage{subcaption}
\usepackage{xspace}

\usepackage{enumitem}
\setlist[itemize]{leftmargin=*}
\setlist[enumerate]{leftmargin=*}

\hypersetup{draft}

\usepackage{algorithm}
\usepackage[noend]{algpseudocode}

\usepackage{listings}
\usepackage{flushend}

\EnableBpAbbreviations

\usepackage{tikz}
\tikzset{
  treenode/.style = {shape=rectangle, rounded corners,
                     draw, align=center,
                     top color=white, bottom color=blue!20},
  root/.style     = {treenode, font=\Large, bottom color=red!30},
  env/.style      = {treenode, font=\ttfamily\normalsize},
  dummy/.style    = {circle,draw}
}

\newsavebox{\fmbox}
\newenvironment{smpage}[1]
{\begin{lrbox}{\fmbox}\begin{minipage}{#1}}
{\end{minipage}\end{lrbox}\usebox{\fmbox}}

\theoremstyle{definition}
\newtheorem{definition}{Definition}
\newtheorem{property}{Property}

\usepackage[frozencache,cachedir=.]{minted}
\input{sections/macro}

\begin{document}

\title{Synthesizing Analytical SQL Queries from Computation Demonstration}

\author{Xiangyu Zhou}
\email{xiangz28@cs.washington.edu}
\affiliation{%
  \institution{University of Washington}
  %\streetaddress{P.O. Box 1212}
  \city{Seattle}
  \state{WA}
  \country{USA}
}
\author{Rastislav Bodik}
\email{bodik@cs.washington.edu}
\affiliation{%
  \institution{University of Washington}
  %\streetaddress{P.O. Box 1212}
  \city{Seattle}
  \state{WA}
  \country{USA}
}
\author{Alvin Cheung}
\email{akcheung@cs.berkeley.edu}
\affiliation{%
  \institution{University of California, Berkeley}
  %\streetaddress{P.O. Box 1212}
  \city{Berkeley}
  \state{CA}
  \country{USA}
}
\author{Chenglong Wang}
\email{chenglong.wang@microsoft.com}
\affiliation{%
  \institution{Microsoft Research}
  \city{Redmond}
  \state{WA}
  \country{USA}
}

\begin{abstract}
Analytical SQL is widely used in modern database applications and data analysis. 
However, its partitioning and grouping operators are challenging for novice users. Unfortunately, programming by example, shown effective on standard SQL, are less attractive because  examples for analytical queries are more laborious to solve by hand. 

To make demonstrations easier to create, we designed a new end-user specification, \emph{programming by computation demonstration}, that allows the user to demonstrate the task using a (possibly incomplete) cell-level computation trace. 
This specification is exploited in a new abstraction-based synthesis algorithm to prove that a partially formed query cannot be completed to satisfy the specification, allowing us to prune the search space. 

We implemented our approach in a tool named \tool and tested it on 80 real-world analytical SQL tasks. Results show that even from small demonstrations, \tool can solve 76 tasks, in 12.8 seconds on average, while the prior approaches can solve only 60 tasks and are on average 22.5$\times$ slower. Our user study with 13 participants reveals that our specification increases user efficiency and confidence on challenging tasks.
\end{abstract}

\maketitle

\input{sections/introduction}
\input{sections/overview}

\input{sections/problem}
\input{sections/algorithm}
\input{sections/experiments}

\input{sections/related}

\section{Conclusion}

We proposed a new synthesis-based approach to make analytical SQL more accessible. We designed (1) a new synthesis specification, \emph{programming by computation demonstration}, that allows the user to demonstrate the task using partial cell-level computation trace, and (2) a new abstraction-based algorithm that leverages \emph{abstract data provenance} to prune infeasible search space more effectively. We implemented and tested our approach on 80 real-world analytical SQL tasks. Results show that \tool can efficiently solve 76 tasks from small demonstrations with a 600 second timeout. In the future, we envision \tool can work with novel user interaction models and ranking strategies to make demonstration creation and solution disambiguation easier.

%
% The acknowledgments section is defined using the "acks" environment
% (and NOT an unnumbered section). This ensures the proper
% identification of the section in the article metadata, and the
% consistent spelling of the heading.
\begin{acks}
This work is supported in part by NSF grants ITE--2132318, ITE--2029457, ITE--1936731, CCF--1918027, IIS--1924435, IIS-1955488, IIS-2027575, CCF-1723352, ARO W911NF2110339, ONR N00014-21-1-2724; 
the Intel and NSF joint research center for Computer Assisted Programming for Heterogeneous Architectures (CAPA NSF CCF-1723352); the CONIX Research Center, one of six centers in JUMP, a Semiconductor Research Corporation (SRC) program sponsored by DARPA CMU 1042741-394324 AM01; DOE award DE-SC0016260; DARPA FA8750–16–2–0032; gifts from Adobe, Google, Intel, Qualcomm, the Sloan Foundation, Adobe, and Facebook. We would also like to thank anonymous reviewers for their insightful feedback.
\end{acks}

\bibliography{main}

%\appendix

\end{document}

%% file: sections/macro.tex
\newcommand{\tool}{\textsc{Sickle}\xspace}
\newcommand{\eval}[1]{{[\![#1]\!]}}
\newcommand{\code}[1]{\mintinline[breaklines,fontsize=\small,escapeinside=||]{postgresql}{#1}}

\newcommand\concat{+\kern-1.3ex+\kern0.8ex}

\newif\ifcomments
\commentstrue
\ifcomments
    \providecommand{\alvin}[1]{{\protect\color{purple}{\bf [alvin: #1]}}}
\else
    \providecommand{\alvin}[1]{}
\fi

%% file: sections/introduction.tex
%!TEX root=../main.tex

\section{Introduction}
\label{sec:intro}

While standard SQL is the de facto language for \emph{data retrieval tasks}, the language of choice for database applications, wrangling routines and other \emph{computation tasks} is analytical SQL~\cite{chaudhuri1997overview,2011-data-wrangling,2012-enterprise-analysis-interviews}. While retrieval queries tend to use operators like select, join and filter, analytical queries see more frequent use of group-aggregation, custom arithmetic functions, and subqueries. Analytical SQL is significantly more powerful than SQL, thanks to its support of the critical ``\code{Partition By}'' operator, which can express partition-aggregation tasks such as the moving average, window aggregation, and ranking.

The combination of partitioning and aggregation computes values over a group of rows and returns a result for \emph{each row}. This differs from aggregation queries in SQL (constructed from the ``\code{Group By}'' operator and aggregation functions), which returns only a single result for \emph{each group of rows}. 
For example, given the input table $T$ below, the aggregation query $q_1$ that sums the sales value for each product ID returns the table $T_1$ with two rows. In contrast, the analytical query $q_2$ partitions the table in place and calculates aggregated values in a new column. % next to original ones. 
%This seemingly subtle difference makes analytical SQL significantly more expressive, and it can solve many tasks that are difficult or unachievable in SQL: besides the \code{CumSum} example shown above, we can use the \code{Max} function with the partition function to solve the ``arg-max'' problem (e.g., find the months with highest sales for each product), or use the \code{Rank} function to rank months by their sales. 
Analytical SQL's increased expressiveness make it harder to program, especially for inexperienced engineers and data scientists~\cite{parameswaran2019enabling}.

%\begin{center}

%   \begin{smpage}{\linewidth}
  
%     \noindent$q_1:$~\mintinline[breaklines,fontsize=\small,escapeinside=||]{postgresql}{Select ID, Sum(Sales) From T Group By ID;}
    
%     \noindent$q_2:$~\mintinline[breaklines,fontsize=\small,escapeinside=||]{postgresql}{Select ID, Quarter, CumSum(Sales) Over (Partition By ID) From T;}
%     \bigskip
%     \end{smpage}
%     \begin{smpage}{\linewidth}{
%       \renewcommand{\arraystretch}{0.9}
%     \footnotesize
%       \begin{tabular}{|ccc|}
%       \multicolumn{3}{l}{\small $T$}\\
%       \hline
%       ID & Quarter & Sales\\\hline
%       \rowcolor{red!7}A  & 1  & 10 \\
%       \rowcolor{red!7}A  & 2  & 20 \\
%       \rowcolor{red!7}A  & 3  & 15 \\
%       \rowcolor{blue!7}B  & 1  & 20 \\
%       \rowcolor{blue!7}B  & 2  & 15 \\\hline
%     \end{tabular}\quad
%     \begin{tabular}{|cc|}
%       \multicolumn{2}{l}{\small $T_1=q_1(T)$}\\
%       \hline
%       ID & Sum\\\hline
%       \rowcolor{red!7}A  & 45 \\
%       \rowcolor{blue!7}B  & 35 \\\hline
%     \end{tabular}\quad
%     \begin{tabular}{|ccc|}
%       \multicolumn{3}{l}{\small $T_2=q_2(T)$}\\
%       \hline
%       ID & Quarter & CumSum\\\hline
%       \rowcolor{red!7}A  & 1  & 10\\
%       \rowcolor{red!7}A  & 2   & 30\\
%       \rowcolor{red!7}A  & 3   & 45\\
%       \rowcolor{blue!7}B  & 1   & 20\\
%       \rowcolor{blue!7} B  & 2  & 35\\\hline
%     \end{tabular}}
%   \end{smpage}
%\end{center}

\begin{smpage}{0.95\linewidth}
\vspace{10pt}
\noindent$q_1:$~\mintinline[breaklines,fontsize=\small,escapeinside=||]{postgresql}{Select ID, Sum(Sales) From T Group By ID}

\noindent$q_2:$~\mintinline[breaklines,fontsize=\small,escapeinside=||]{postgresql}{Select ID, Quarter, CumSum(Sales)
Over (Partition By ID) From T;}
\end{smpage}
\vspace{-3pt}
\begin{center}
  \begin{smpage}{0.86\linewidth}
  \begin{tikzpicture}
    [
      grow                    = right,
      sibling distance        = 5em,
      level distance          = 11em,
      edge from parent/.style = {draw, -latex},
      every node/.style       = {font=\scriptsize}  
    ]
    \node {
      \begin{tabular}{|ccc|}
  \multicolumn{3}{l}{\small $T$}\\
  \hline
  ID & Quarter & Sales\\\hline
  \rowcolor{red!7}A  & 1  & 10 \\
  \rowcolor{red!7}A  & 2  & 20 \\
  \rowcolor{red!7}A  & 3  & 15 \\
  \rowcolor{blue!7}B  & 1  & 20 \\
  \rowcolor{blue!7}B  & 2  & 15 \\\hline
  \end{tabular}}
      child { node  {
      \begin{tabular}{|ccc|}
          \multicolumn{3}{l}{\small $T_2$}\\
          \hline
          ID & Quarter & CumSum\\\hline
          \rowcolor{red!7}A  & 1  & 10\\
          \rowcolor{red!7}A  & 2   & 30\\
          \rowcolor{red!7}A  & 3   & 45\\
          \rowcolor{blue!7}B  & 1   & 20\\
         \rowcolor{blue!7} B  & 2  & 35\\\hline
          \end{tabular}
  }
        edge from parent node [above] {$q_2$} }
      child { node {
          \begin{tabular}{|cc|}
              \multicolumn{2}{l}{\small $T_1$}\\
              \hline
              ID & Sum\\\hline
              \rowcolor{red!7}A  & 45 \\
              \rowcolor{blue!7}B  & 35 \\\hline
              \end{tabular}
      }
        edge from parent node [above] {$q_1$} };
  \end{tikzpicture}
  \end{smpage}
\end{center}

Recently, program synthesizers, especially programming-by-example (PBE) tools, have been successfully adopted to solve similar programming challenges in domains such as string manipulation~\cite{DBLP:conf/popl/Gulwani11,polozov2015flashmeta}, SQL/Datalog query~\cite{DBLP:journals/pacmpl/RaghothamanMZNS20,DBLP:conf/pldi/WangCB17,DBLP:conf/pldi/ThakkarNSANR21,DBLP:conf/sigsoft/SiLZAKN18}, data wrangling~\cite{DBLP:journals/pacmpl/WangFBCD20,DBLP:conf/pldi/FengMBD18,DBLP:conf/pldi/FengMGDC17}. 
In such tools, the user provides one or more pairs of small input-output example values $(I,O)$; the tool then synthesizes a program (or a list of ranked programs) $p$ that satisfies $\eval{p(I)}=O$. 
%This lets inexperienced users solve the task without programming. 
Though promising, PBE is less effective on analytical tasks for two reasons:%due to their limitations to address the user specification and the algorithm design challenges introduced by analytical SQL --- the need to handle
% as they cannot effectively express %fail to synthesize 
% complex computations in analytical SQL that aggregate and derive values. We summarize the reasons as follows:

%\paragraph{The User Specification Challenge} While I/O examples can succinctly capture the user intent in many domains, they are not ideal for analytical tasks:

%SQL queries with aggregations~\cite{DBLP:conf/chi/WangFBDCK21} due to the need to compute aggregation values. With the introduction of more complex computations, creating examples to illustrate analytical SQL tasks is even harder. Concretely, there are three specification challenges:
\begin{itemize}
% \item \textbf{Example size:} Because analytical functions operate in groups of rows instead of individual rows (as in SQL or Datalog), a representative example often contain multiple groups of rows. %and a small example can be too ambiguous to capture the task intent. 
\item \textbf{Increased specification effort:} 
% As analytical functions operate on groups of rows instead of individual rows (as in SQL or Datalog), a representative example often contains results computed from multiple groups of rows. Providing an I/O example requires users to manually find all the values that belong to a specific group and compute the aggregated value from them.
%compute the aggregated values from groups of input values (which requires finding all of the values belong to the same group and summing them up, for all groups).
% This is costly and often results in erroneous I/O  examples (as reported in~\cite{DBLP:conf/chi/WangFBDCK21}).
%
As analytical functions operate on partitions of rows, a representative example often contains results computed from multiple groups of rows. Providing an example requires users to manually find all the values that belong to a specific group and compute the aggregated value from them.
%compute the aggregated values from groups of input values (which requires finding all of the values belong to the same group and summing them up, for all groups).
This is costly when groups are created by partitioning and often results in erroneous examples (as reported in~\cite{DBLP:conf/chi/WangFBDCK21}).
% \item \textbf{Numerical Issue:} Analytical queries frequently use aggregation and arithmetic functions that involve floating-point value computations (e.g., moving average, percentage). Thus, the user output examples can be numerically imprecise, which make correctness checking more challenging. This problem was often omitted, or it needs a more expensive approximate synthesis algorithm to handle~\cite{DBLP:conf/sigsoft/HandaR20,DBLP:conf/ecoop/PelegP19}.
\item \textbf{Limited synthesis efficiency:} Because examples capture only output values, the synthesizer needs to ``guess'' the functions used in the computation. This makes reasoning about analytical queries much harder and makes the synthesizer difficult to scale up, especially when custom arithmetic functions are needed.
%This makes examples more ambiguous% 
% and affects the synthesis speed
%, especially when custom arithmetic functions are needed.
\end{itemize}

\noindent In this paper, we propose 
%our first contribution is a new synthesis specification design, 
\emph{programming by computation demonstration} to solve the specification challenge. Our design is motivated by the observation that online users often leverage example formulas to explain tasks. In our new specification:
\begin{enumerate}
  \item The user provides the input tables and a \emph{partial} output table to demonstrate the task.
  \item In the partial output table, the user provides expressions (similar to Excel formulas) to demonstrate how output values are computed from values in the inputs, as opposed to providing only the final values.
  \item The user can optionally provide \emph{incomplete expressions} when the expression involves many values to further reduce specification effort.
\end{enumerate}
\noindent We conducted a user study with 13 participants on 6 real-world scenarios and compared user experiences using computation demonstration versus classical PBE (\autoref{sec:experiments}). 
Our quantitative results show that both approaches have similar user efficiencies on simpler tasks, but our method made users more efficient on harder tasks. Our qualitative results furthermore show that users are more confident using computation demonstration and generally prefer it over examples. %due to its expressiveness. %(1) user performances have no significant differences on simpler tasks using two specifications, and (2)

% because the demonstration contains the computation process, it exposes custom functions involved in the computation and avoids numerical issues, making the specification less ambiguous and more robust. For example, the user can use a partial table {\scriptsize\begin{tabular}{|cccc|}
% \hline c1 & c2 & c3 & c4\\
% \hline A  & 1 & 10  & $\mathsf{sum}(10)$ \\
% A  & 2  & 20 & $\mathsf{sum}(10, \diamondsuit)$ \\\hline\end{tabular}}
% to demonstrate the cumulative sum task in \autoref{fig:analytical-sql-examples}-2, where each value is a reference to an input table cell and ``$\diamondsuit$'' stands for omitted values.

Because a computation demonstration constrains the structure of the desired computation but does not provide the output value, the classical PBE objective ``$\eval{p(I)}=O$'' no longer characterizes our synthesis problem. We develop a new synthesis task formulation based on \emph{provenance consistency}, which is a property of a program whose computation traces generalize the user demonstration. This formulation builds on a new query semantics (\autoref{sec:problem}) called the \emph{provenance-tracking query semantics} that keeps track of cell-level data provenance during evaluation.

%\paragraph{The Algorithm Design Challenge} 
Our new specification raises new challenges and opportunities for algorithm design. A promising approach for our synthesis task is abstraction-based search approach that has been successfully adopted in synthesis of list and table transformation programs~\cite{DBLP:conf/pldi/FengMBD18,DBLP:journals/pacmpl/MiltnerFPWZ18,DBLP:journals/pacmpl/WangFBCD20,DBLP:conf/pldi/WangCB17}: the synthesizer iteratively enumerates and prunes partial programs until the correct solution is found. The key in this approach is to leverage an abstract interpreter to reason about the realizability of the synthesis goal given a partial program, and the synthesizer's scalability depends on whether the abstraction can effectively capture inconsistent behavior of incorrect partial programs for early pruning. However, existing abstractions are insufficient for analytical SQL (as shown in our experiments~\autoref{sec:experiments}): abstractions based on high-level type properties of full outputs (row/column/group numbers)~\cite{DBLP:conf/pldi/FengMBD18,DBLP:journals/pacmpl/MiltnerFPWZ18} are not ideal for partial demonstration whose type information is incomplete~\cite{DBLP:journals/pacmpl/WangFBCD20}; value-based abstractions that reason program properties by tracking concrete value flow~\cite{DBLP:journals/pacmpl/WangFBCD20,DBLP:conf/pldi/WangCB17} are also insufficient, as analytical operators can mutate values -- making them intractable by current abstractions; and computation graph based abstractions~\cite{DBLP:journals/pacmpl/BavishiLSS21} are inapplicable due to the user's computation demonstration contains partial expressions.

Hence, our second contribution is a new abstract semantics, \emph{abstract data provenance}, developed to better capture the semantics of partial queries with analytical computations. Our key insight is to leverage the fine-grained provenance information presented by computation demonstrations to prune partial queries that cannot be completed. Given a partial query, the abstract analyzer over-approximates the provenances of each output cell and checks their consistency with the user demonstration. This analysis reveals cell-level provenance inconsistency introduced by infeasible queries that other abstractions cannot capture. The new abstraction lets our algorithm on average visit 97.08\% less queries.
%: e.g., a cell in $\mathcal{E}$ may need to use two input values from $T$ but these two values cannot reach the same output cell according to the analysis (due to incorrect partitioning parameters). 
%Because of its better use of the user demonstration, the provenance abstraction achieves better precision and leads to better performance (on average visit 85.06\% less programs because of its superior pruning power).

%\medskip

We implemented our algorithm in a tool called \tool and evaluated it on {80} tasks: {60} tasks from online posts and {20} from the popular TPC-DS~\cite{DBLP:conf/vldb/OthayothP06} database benchmark. Our experiments show that \tool can solve 76 of these benchmarks and outperforms prior state-of-the-art techniques~\cite{DBLP:journals/pacmpl/WangFBCD20,DBLP:conf/pldi/FengMBD18} that only solves 60 and 51 benchmarks. On benchmarks all techniques can solve, \tool is on average 22.5$\times$ faster.
% \paragraph{Benefit of demonstration using trace}
% \begin{itemize}
%     \item Easier for user to provide (cite Falx paper)
%     \item A more precise specification for the synthesizer
%     \item Robust against practical issue like floating point number
% \end{itemize}

% \paragraph{Contribution of the synthesis algorithm}
% \begin{itemize}
%     \item comparing to Morpheus abstraction, it provides richer information (baseline)
%     \item comparing to Scythe abstraction, the trace storage is more concise (factored representation) and thus scales better to large inputs
% \end{itemize}

%\alvin{cut the following for space}
\paragraph{Contributions} This paper's contributions include:
\begin{itemize}
    \item We introduce the analytical SQL query synthesis problem and present a new user specification, \emph{computation demonstration}, that makes specifying analytical tasks easier.
    \item We conducted a user study to compare user experiences in creating computation demonstrations and examples.
%    \item We formalized the computation consistent program synthesis task with the introduction of provenance tracking query semantics.
    \item We propose an abstraction-based algorithm with a new language abstraction, \emph{abstract data provenance}, that can better utilize finer-grained computation information to dramatically prune infeasible programs.
    \item We implemented our approach as a practical tool, \tool, and evaluated it on 80 real-world benchmarks. Results show that \tool can efficiently solve practical analytical SQL tasks with much better performance comparing against prior state-of-the-art algorithms. 
\end{itemize}

%% file: sections/overview.tex
%!TEX root=../main.tex

\begin{figure*}[t]
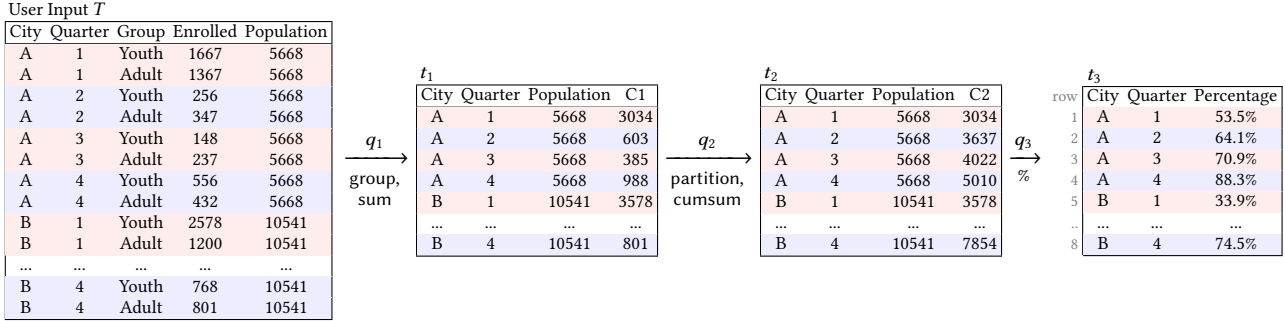

	{
	\setlength\tabcolsep{1.5pt}
	{\scriptsize
	\begin{tabular}{|ccccc|}
	\multicolumn{5}{l}{User Input $T$}\\
	\hline
	City	& Quarter & Group  & Enrolled & Population \\\hline
	\rowcolor{red!7}A		&   1  & Youth  & 1667     & 5668 \\
	\rowcolor{red!7}A   	&   1  & Adult  & 1367     & 5668 \\
	\rowcolor{blue!7}A		&   2  & Youth  & 256      & 5668 \\
	\rowcolor{blue!7}A    	&   2  & Adult  & 347      & 5668 \\
	\rowcolor{red!7}A 		&   3  & Youth  & 148      & 5668 \\
	\rowcolor{red!7}A    	&   3  & Adult  & 237      & 5668 \\
	\rowcolor{blue!7}A    	&   4  & Youth  & 556      & 5668 \\
	\rowcolor{blue!7}A    	&   4  & Adult  & 432      & 5668 \\
	\rowcolor{red!7}B    	&   1  & Youth  & 2578     & 10541\\
	\rowcolor{red!7}B    	&   1  & Adult  & 1200     & 10541\\
	...     & ...  & ...    & ...      & ... \\
	\rowcolor{blue!7}B    	&   4  & Youth  & 768      & 10541\\
	\rowcolor{blue!7}B    	&   4  & Adult  & 801      & 10541\\\hline
	\end{tabular}
	}
	$\xrightarrow[{\setlength\arraycolsep{0pt}\scriptsize\begin{array}{c}\mathsf{group,}\\\mathsf{sum}\end{array}}]{q_1}$
	{\scriptsize\begin{tabular}{|cccc|}
	\multicolumn{4}{l}{$t_1$}\\
	\hline
	City	& Quarter & Population & C1\\\hline
	\rowcolor{red!7}A		&   1     & 5668 & 3034\\
	\rowcolor{blue!7}A    	&   2  	  & 5668 &  603\\
	\rowcolor{red!7}A    	&   3  	  & 5668 &  385\\
	\rowcolor{blue!7}A    	&   4  	  & 5668  &  988\\
	\rowcolor{red!7}B    	&   1     & 10541 &  3578\\
	... & ... & ...  & ... \\
	\rowcolor{blue!7}B    	&   4     & 10541 & 801     \\\hline
	\end{tabular}}
	$\xrightarrow[{\setlength\arraycolsep{0pt}\scriptsize\begin{array}{c}\mathsf{partition,}\\\mathsf{cumsum}\end{array}}]{q_2}$
	{\scriptsize
	\begin{tabular}{|cccc|}
	\multicolumn{4}{l}{$t_2$}\\
	\hline
	City	& Quarter & Population & C2 \\\hline
	\rowcolor{red!7}A		&   1     & 5668  & 3034      \\
	\rowcolor{blue!7}A    	&   2  	  & 5668  &  3637      \\
	\rowcolor{red!7}A    	&   3  	  & 5668  &  4022      \\
	\rowcolor{blue!7}A    	&   4  	  & 5668  &  5010     \\
	\rowcolor{red!7}B    	&   1     & 10541 &  3578\\
	... & ... & ...  & ... \\
	\rowcolor{blue!7}B    	&   4     & 10541 &  7854\\\hline
	\end{tabular}}
	$\xrightarrow[{\scriptsize \%}]{q_3}$
	{\scriptsize\begin{tabular}{r|ccc|}
	\multicolumn{1}{l}{} &\multicolumn{3}{l}{$t_3$}\\
	\cline{2-4}
	{\color{gray}\tiny row} & City	& Quarter  & Percentage \\\cline{2-4}
	\rowcolor{red!7} \cellcolor{white}{\color{gray}\tiny 1} & A		&   1   & 53.5\%     \\
	\rowcolor{blue!7} \cellcolor{white}{\color{gray}\tiny 2} &A		&   2   & 64.1\%      \\
	\rowcolor{red!7} \cellcolor{white}{\color{gray}\tiny 3} &A 		&   3   & 70.9\%     \\
	\rowcolor{blue!7} \cellcolor{white}{\color{gray}\tiny 4} &A    	&   4   & 88.3\%      \\
	\rowcolor{red!7} \cellcolor{white}{\color{gray}\tiny 5} &B    	&   1   & 33.9\%    \\
	{\color{gray}\tiny ..} & ... & ... & ... \\
	\rowcolor{blue!7} \cellcolor{white}{\color{gray}\tiny 8} & B    	&   4   & 74.5\%\\\cline{2-4}
	\end{tabular}}
	}
	\caption{The running example: given the input table $T$, the user aims to calculate the percentage of total population that have enrolled in the program at the end of the each quarter (for each city). The solution requires three steps ($q$ in \autoref{fig:running-example}).}
	% The solution is shown in \protect\autoref{fig:running-example}\protect, and $q_1, q_2, q_3$ are the three steps needed to compute the final result $t_3$.}
	\label{fig:overview-running-example-big}
	\end{figure*}

\section{Overview}
\label{sec:overview}

In this section, we use an example to illustrate how a user specifies an analytical task using computation demonstration, and how \tool solves the problem.

\paragraph{The Task} Suppose the user has an input table $T$ (\autoref{fig:overview-running-example-big}) with the number of people enrolled in a health program split by city, quarter, and age group (the {\small\tt Population} column shows the city population). 
%The table $T$ has five columns {\small\tt City}, {\small\tt Quarter}, {\small\tt Group}, {\small\tt Enrolled}, and {\small\tt Population} (population of the city). 
For each city, the user wants to compute the percentage of total population enrolled in the program at the end of each quarter. This can be done using the analytic SQL query $q$ shown in \autoref{fig:running-example}: (1) the subquery $q_1$ calculates the total enrollment in each city for each quarter
%It groups the table by {\small\tt City}, {\small\tt Quarter} and {\small\tt Population} and summing up the {\small\tt Enrolled} column; 
(2) the subquery $q_2$ then calculates the cumulative sum of the enrollment number for each city at the end of each quarter, by partitioning on \code{City} 
%(so that the cumulative sums are calculated within each city) 
and applying \code{Cumsum} on the column \code{C1} generated by $q_1$;
(3) the subquery $q_3$ finally calculates the enrollment percentage based on \code{C2} and \code{Population}.
%it calculates the percentage of city population enrolled at the end of each quarter for each city 
\autoref{fig:overview-running-example-big} shows the output $t_i$ of each subquery $q_i$.

Because this is a challenging task that requires using analytical function together with subqueries, grouping and arithmetic, the user decides to solve it with \tool.

\begin{figure}[ht]
\begin{minted}[fontsize=\footnotesize,escapeinside=||]{postgresql}
--  q3: calculate percentage
Select City, Quarter, C2 / Population * 100%
From ( 
  --  q2: calculate cumulative enrolled number
  Select City, Quarter, Population,
         Cumsum(C1) Over (Partition By City) As C2
  From (
    -- q1: calculate the number of people enrolled in
    --     the program for each city / quarter
    Select City, Quarter, Population, Sum(Enrolled) As C1
    From T
    Group By City, Quarter, Population));
\end{minted}
\caption{The desired query $q$ to solve the task in \protect\autoref{fig:overview-running-example-big}.}
\label{fig:running-example}
\end{figure}

\subsection{The User Specification}\label{subsec:overview-userspec} To use \tool, the user creates a \emph{computation demonstration} to demonstrate the task. Instead of providing the full output ($t_3$ in \autoref{fig:overview-running-example-big}), the user creates a partial output that shows how output cells are derived from the input. 

\begin{figure}[t]
\includegraphics[width=0.7\linewidth]{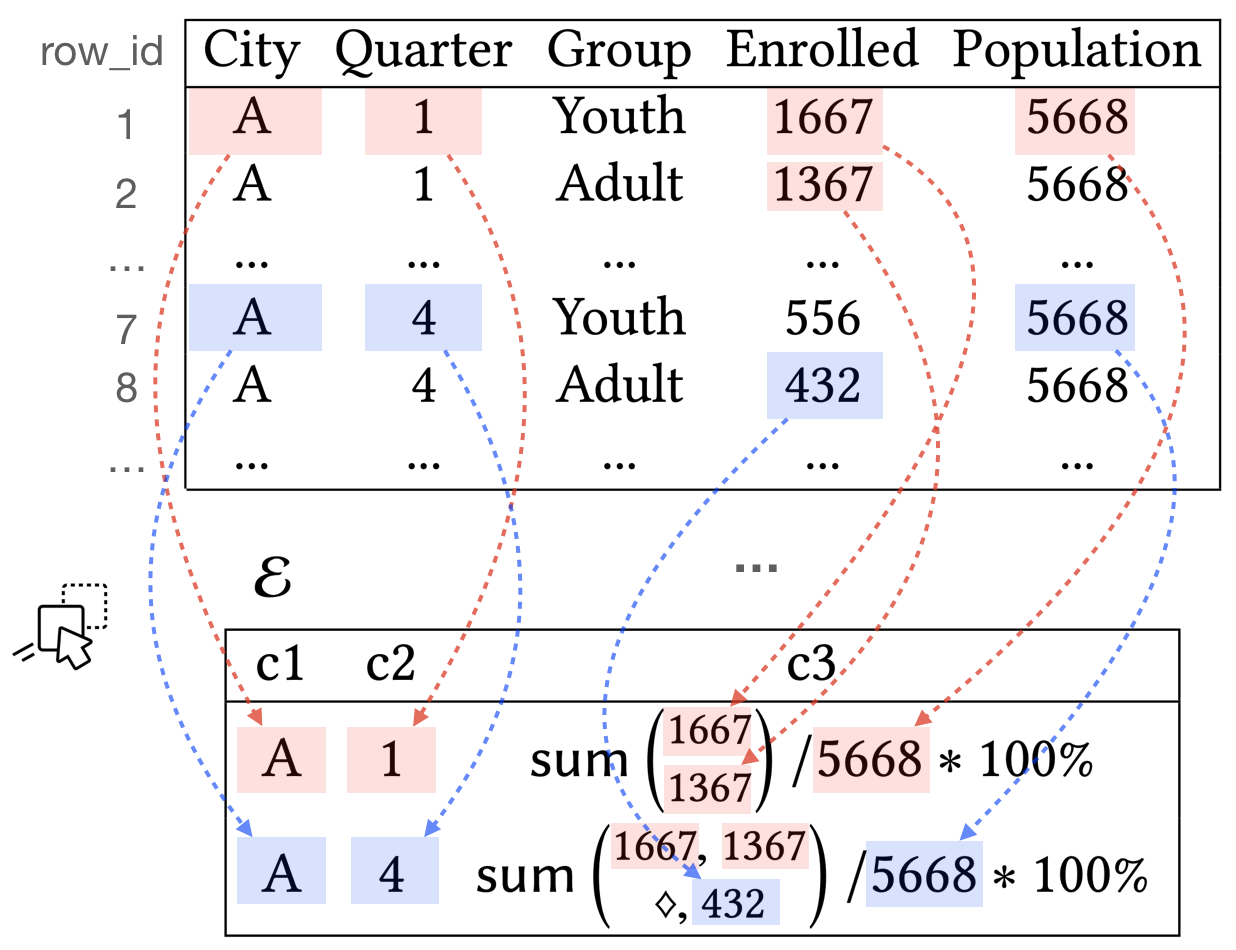}

\medskip

{
\footnotesize\begin{tabular}{|ccc|}
\multicolumn{3}{l}{\color{darkgray}\emph{Internal representation of $\mathcal{E}$:}}\\
\hline
c1	    &  c2  & c3 \\\hline
$T[1,1]$		&   $T[1,2]$       & ${\color{blue}\mathsf{sum}(}T[1, 4], T[2, 4]{\color{blue})} / T[1,5] * 100\%$     \\
$T[7,1]$    	&   $T[7,2]$       & ${\color{blue}\mathsf{sum}(}T[1, 4], T[2, 4], \diamondsuit, T[8, 4]{\color{blue})} / T[7,5] * 100\%$      \\\hline
\end{tabular}}

\caption{The user demonstration $\mathcal{E}$ constructed by dragging and dropping input cells to the partial output table. The symbol ``$\diamondsuit$'' denotes that the user omitted some values in the expression. The table in the bottom shows internal representation of the demonstration (based on references of input cells). $T[i,j]$ denotes the reference to the cell at row $i$ and column $j$ in the input $T$ (index starts from $1$).}
\label{fig:overview-user-demonstration}
\end{figure}

\paragraph{The User Demonstration} \autoref{fig:overview-user-demonstration} shows the user demo $\mathcal{E}$, constructed by expressions and references to input table cells. Here, the user demonstrates how the percentage is computed for quarter 1 and 4 of city A (i.e., the first and fourth row of the output $t_3$ in \autoref{fig:overview-user-demonstration}):
\begin{itemize}
\item The percentage of people enrolled at the end of quarter 1 in city A is computed by (1) summing up the number of people enrolled in both age groups \colorbox{black!7}{$\mathsf{sum}(1667, 1367)$}, and (2) dividing the sum with the city population $5668$.
\item For quarter 4, the percentage requires dividing the number of people from both age groups enrolled throughout quarters 1 to 4 with the population $5668$. As it requires collecting many values from $T$, the user uses an incomplete expression \colorbox{black!7}{$\mathsf{sum}(1667, 1367, \diamondsuit, 432)$} in the demonstration ($\diamondsuit$ denotes omitted values) to save effort, where the enrollment from rows 3-7 of $T$ are omitted.
\end{itemize}

\noindent With a PBE tool, the user would need to manually group values from the input and compute all the derived values ($t_3$ in \autoref{fig:overview-running-example-big}) with considerable effort; but using \tool, the user constructs the demonstration easily by dragging and dropping input cells to create a \emph{partial output}. Additionally, the support for incomplete expressions permits users to not exhaustively find all values needed for the $\mathsf{sum}$ expression, and such support would be especially beneficial if values needed the expression are not in a continuous region in the input data (e.g., if the user wants to compute the percentage of enrolled Youth participants over all enrolled participants in all cities over all four quarters). %By allowing partial output and incomplete expressions, there is less specification effort compared to I/O examples. 
Because the user's drag-and-drop actions provide references to input cells (not just their values), \tool internally stores cell references in $\mathcal{E}$ and uses them in the synthesis process (\autoref{fig:overview-user-demonstration}). %For simplicity, we use the example in \autoref{fig:overview-user-demonstration} throughout the section and discuss the internal representation in \autoref{sec:problem}.

\paragraph{Synthesis Consistency Criteria} With our new specification, we formulate the new \emph{computation consistency} criteria for deciding whether a synthesized query $q$ satisfies the user demonstration. We first introduce a new semantics of analytical SQL: \emph{provenance-tracking query semantics}. Under this semantics, query operators are ``term rewriters'' that transform and simplify cell-level expressions symbolically to keep track of cell-level data provenance.

% {
% \renewcommand{\arraystretch}{1.2}
% \footnotesize\begin{tabular}{|ccc|}
% \hline
% c1	    &  c2  & c3 \\\hline
% $T[1,1]$		&   $T[1,2]$       & $\mathsf{sum}({\setlength\arraycolsep{0pt}\tiny\begin{array}{c}T[1, 4]\\ T[2, 4]\end{array}}) / T[1,5] * 100\%$     \\
% $T[8,1]$    	&   $T[8,2]$       & $\mathsf{sum}({\setlength\arraycolsep{0pt}\tiny\begin{array}{c}T[1, 4], T[2, 4]\\\circ, T[8, 4])\end{array}}) / T[7,5] * 100\%$      \\\hline
% \end{tabular}}

\begin{figure}[h]
	{
	\setlength\tabcolsep{1.6pt}
	{\scriptsize
	\setlength\tabcolsep{2pt}
	\begin{tabular}{r|ccccc|}
	\multicolumn{1}{l}{} & \multicolumn{4}{l}{$T$}\\
	\cline{2-6}
	{\color{gray}\tiny row}& City	& Quarter & Group  & Enrolled & Population \\\cline{2-6}
	\rowcolor{red!7} \cellcolor{white}{\color{gray}\tiny 1}& $T[1,1]$		&   $T[1,2]$  & $T[1,3]$  & $T[1,4]$     & $T[1,5]$ \\
	\rowcolor{red!7} \cellcolor{white}{\color{gray}\tiny 2}& $T[2,1]$   	&   $T[2,2]$  & $T[2,3]$  & $T[2,4]$     & $T[2,5]$ \\
	\cellcolor{white}{\color{gray}\tiny ..} & ...     & ...  & ...    & ...      & ... \\
	\rowcolor{blue!7} \cellcolor{white}{\color{gray}\tiny 7}& $T[7,1]$    	&   $T[7,2]$  & $T[7,3]$  & $T[7,4]$      & $T[7,5]$ \\
	\rowcolor{blue!7} \cellcolor{white}{\color{gray}\tiny 8}& $T[8,1]$    	&   $T[8,2]$  & $T[8,3]$  & $T[8,4]$      & $T[8,5]$ \\
	\cellcolor{white}{\color{gray}\tiny ..}     & ...  & ...    & ...      & ... & ... \\\cline{2-6}
	\end{tabular}
	}
	$\xrightarrow[{\tiny \mathsf{}}]{q}$
	{\footnotesize\begin{tabular}{r|ccc|}
	 \multicolumn{1}{l}{} & \multicolumn{3}{l}{$t_3^\star$\vspace{2pt}}\\
	\cline{2-4}
	{\color{gray}\tiny row} & City	& Quarter & Percentage\\\cline{2-4}
	\rowcolor{red!7} \cellcolor{white}{\color{gray}\tiny 1} & $\mathsf{group}\{{\setlength\arraycolsep{0pt}\tiny\begin{array}{c}T[1,1]\\T[2,1]\end{array}}\}$		&  $\mathsf{group}\{{\setlength\arraycolsep{0pt}\tiny\begin{array}{c}T[1,2]\\T[2,2]\end{array}}\}$  & $\mathsf{sum}({\setlength\arraycolsep{0pt}\tiny\begin{array}{c}T[1,4]\\T[2,4]\end{array}}) / \mathsf{group}\{{\setlength\arraycolsep{0pt}\tiny\begin{array}{c}T[1,5]\\T[2,5]\end{array}}\} * 100\%$\\
	{\color{gray}\tiny ..} & ... & ... & ...   \\
	\rowcolor{blue!7} \cellcolor{white}{\color{gray}\tiny 4} & $\mathsf{group}\{{\setlength\arraycolsep{0pt}\tiny\begin{array}{c}T[7,1]\\T[8,1]\end{array}}\}$		&   $\mathsf{group}\{{\setlength\arraycolsep{0pt}\tiny\begin{array}{c}T[7,2]\\T[8,2]\end{array}}\}$  & {\tiny ${\mathsf{sum}\left({\setlength\arraycolsep{0pt}\tiny\begin{array}{c}T[1,4], T[2,4] \\T[3,4],T[4,4]\\T[5,4],T[6,4]\\ T[7,4], T[8,4]\end{array}}\right)} / \mathsf{group}\{{\setlength\arraycolsep{0pt}\tiny\begin{array}{c}T[7,5]\\T[8,5]\end{array}}\} * 100\%$}\\
	{\color{gray}\tiny ..}& ... & ... & ...   \\\cline{2-4}
	\end{tabular}}
	}
	\caption{The evaluation result of $q$ (\autoref{fig:running-example}) on the input $T$ (from \autoref{fig:overview-running-example-big}) under provenance tracking semantics. The output $t_3^\star$ shows how $t_3$'s cells (\autoref{fig:overview-running-example-big}) are derived from $T$.}
	%$T[i,j]$ refers to the cell at $i$-th row and $j$-th column in $T$.}
	\label{fig:overview-provenance-tracking-eval}
\end{figure}

\autoref{fig:overview-provenance-tracking-eval} shows the provenance-tracking evaluation result for the example in \autoref{fig:overview-running-example-big}. For example, the first row of the \code{City} column in $t_3^\star$ is derived by the \code{Group By} operator in the subquery $q_1$; since the \code{Group By} uses both $T[1,1]$ and $T[2,1]$ in the same group, the cell content is \colorbox{red!7}{\small $\mathsf{group}\{T[1,1],T[1,2]\}$}. The percentage in row 4 is obtained by (1) summing up the enrolled number from row 1 to row 8 in $T$, and (2) applying $\lambda x,y.x/y*100\%$ to the sum and the city population.

We now define the synthesis objective based on the following provenance consistency criteria: we consider a query consistent with the user demonstration if its provenance tracking evaluation output $t_0^\star$ satisfies the following criteria:
\begin{itemize}
\item There exists a subtable $t_s^\star$ of $t^\star$ such that every cell in $t_s^\star$ generalizes each cell in the user demonstration $\mathcal{E}$ based on their provenance information. 
\end{itemize}

\noindent In our example, the query $q$ in \autoref{sec:overview} is consistent with $\mathcal{E}$ (\autoref{fig:overview-user-demonstration}): %because its provenance tracking evaluation output $t_3^\star$ (\autoref{fig:overview-provenance-tracking-eval}) satisfies the criteria.%(we denote this generalization criteria as $\mathcal{E}\prec t_3^\star$):
the subtable $t_s^\star$ of $t_3^\star$ (provenance tracking evaluation output of $q$) that witnesses the property is the table consists of only rows 1 and 4 of $t_3^\star$. The generalization is shown by: the term \colorbox{red!7}{\footnotesize $T[1,1]$} (from row 1, column $c_1$ in $\mathcal{E}$) is subsumed by the term \colorbox{red!7}{\footnotesize $\mathsf{group}\{T[1,1], T[2,1]\}$} (row 1, column \code{City} in $t_3^\star$)~\footnote{Since all of the values in the same group-by column have the same value, (e.g., ``$A$'' in this case); the user can use either $T[1,1]$ or $T[1,2]$ in the demonstration of the cell (or both), this makes user demonstration flexible.}; the term \colorbox{blue!7}{\scriptsize$\mathsf{sum}\left({\setlength\arraycolsep{0pt}\tiny\begin{array}{c}T[1, 4], T[2, 4]\\ \diamondsuit, T[8, 4]\end{array}}\right) / T[7,5] * 100\%$} in $\mathcal{E}$ is subsumed by \colorbox{blue!7}{\tiny ${\mathsf{sum}\left({\setlength\arraycolsep{0pt}\tiny\begin{array}{c}T[1,4], T[2,4], T[3,4],T[4,4]\\T[5,4],T[6,4], T[7,4], T[8,4]\end{array}}\right)/ \mathsf{group}\{{\setlength\arraycolsep{0pt}\tiny\begin{array}{c}T[7,5]\\T[8,5]\end{array}}\} * 100\%}$} in $t_3^\star$ (because the user's omitted values denoted by ``$\diamondsuit$'' can match any number of other values).

Now, our goal is to synthesize the computation-consistent query $q$ from the input $T$ and the user demonstration $\mathcal{E}$.

\subsection{The Synthesis algorithm}

To solve the synthesis problem, our algorithm adopts an abstraction-based enumerative search algorithm.
%: starting from query skeletons, the synthesizer iteratively instantiates and reasons about partial queries based on their abstract semantics (and prune infeasible ones) until correct queries are found. 
For simplicity, we consider only three query operators here: $\mathsf{group}$ (grouping and aggregation), $\mathsf{partition}$ (partition and aggregation), and $\mathsf{arithmetic}$. We also represent queries in the following ``instruction'' style. The query $q$ in \autoref{fig:running-example} is shown below (the instruction at line $i$ corresponds to the subquery $q_i$, and $t_i$ is to the output of $q_i$):

\medskip
\begin{center}
\begin{smpage}{\linewidth}
\begin{minted}[fontsize=\footnotesize,escapeinside=||]{R}
t1 <- group(T, [City, Quarter, Population], sum, Enrolled)
t2 <- partition(t1, [City], cumsum, C1)
t3 <- arithmetic(t2, |$\lambda x,y.x/y*100\%$|, [C2, Population])
\end{minted}
\end{smpage}
\end{center}
\medskip

% \medskip
% \begin{center}
% \begin{smpage}{0.7\linewidth}
% \begin{minted}[fontsize=\small,escapeinside=||]{R}

% t1 <- group(T, [City, Quarter, Population], |$\square$|, |$\square$|)
% t2 <- arithmetic(t1, |$\square$|, |$\square$|)
% \end{minted}
% \end{smpage}
% \end{center}
% \medskip

% $$t^\circ_2 = \eval{q_B(T)}^\circ$$

\begin{figure*}[t]
\includegraphics[width=\linewidth]{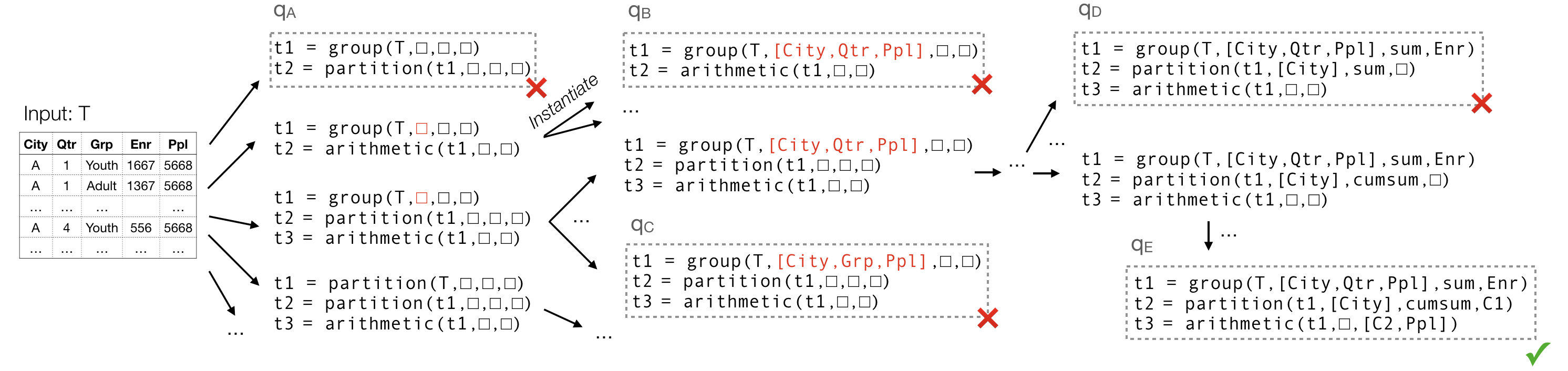}
\caption{The enumerative search algorithm to solve the running example ($\mathcal{E}$ from \autoref{fig:overview-user-demonstration}). Queries in dashed boxes are leaf nodes in the search tree. Infeasible partial programs are pruned based on abstract reasoning.}
\vspace{-5pt}
\label{fig:overview-enumerative-search}
\end{figure*}

\paragraph{Enumerative Search} \autoref{fig:overview-enumerative-search} shows the enumerative search process to solve the running example. From the input table $T$, \tool first enumerates query skeletons formed by compositions of query operators with no instantiated parameters (represented as holes ``$\square$'') and then iteratively instantiates their parameters (e.g., the query $q_C$ is generated by instantiating the first hole of its parent). The search tree expands with the breadth-first-search strategy.

During the search process, an abstract analyzer analyzes whether a partial query can realize the synthesis goal and prunes infeasible ones. The synthesizer terminates when it finds $N$ (a synthesizer parameter, \tool uses $N=10$) consistent solutions or timeout. The search space can be very large in practice (e.g., the search space for the running example contains 1,181,224 queries even only queries up to size 3 are considered), especially because the candidate numbers for many query parameters (e.g., grouping columns) grow exponentially with the input column number. Thus, \tool's performance depends on whether the abstract analyzer can prune infeasible partial queries early. %For a simple program like $q_A$, we can easily prune it based on the fact that: without an $\mathsf{arithmetic}$ function, the output would not contain a term with percentage function. But for other programs, we need stronger program reasoning techniques to analyze about their feasibility, which is achieved by the abstract provenance analyzer we will introduce next.

\paragraph{Pruning with Abstract Provenance Analysis} Our abstraction design is based on the following key observation: incorrect analytical queries are often resulted by wrong partitioning or grouping of columns, including (1) values that are supposed to be aggregated into the same group are spread in different groups, and (2) values supposedly to appear in different groups in the demonstration are wrongly aggregated together. Because recognizing such inconsistency requires fine-grained cell level information tracking, existing abstractions (e.g., type abstractions~\cite{DBLP:conf/pldi/FengMGDC17,DBLP:conf/pldi/PolikarpovaKS16}, value abstractions~\cite{DBLP:conf/pldi/WangCB17}) are insufficient for effective pruning.

We introduce a new abstraction, \emph{abstract data provenance}, that over-approximates cell-level data provenance to keep track of how input cells will be partitioned / grouped / aggregated throughout the computation process for more effective pruning. Given a partial query $q$ (with holes), the abstract analyzer returns a table $t^\circ=\eval{q(T)}^\circ$ (denotes evaluating $q$ on $T$ using abstract semantics). Each cell $t^\circ[i,j]$ is a set of input cells that contains all possible input values that can possibly flow into the position $i,j$. This is an over-approximation of data provenance: for any possible instantiations of $q$, its output cell at $[i,j]$ may only use a subset of input cells from $t^\circ[i,j]$. The analyzer then checks if the provenance information from the user example $\mathcal{E}$ is consistent with $t^\circ$.

\begin{figure*}[t]
\includegraphics[width=0.84\linewidth]{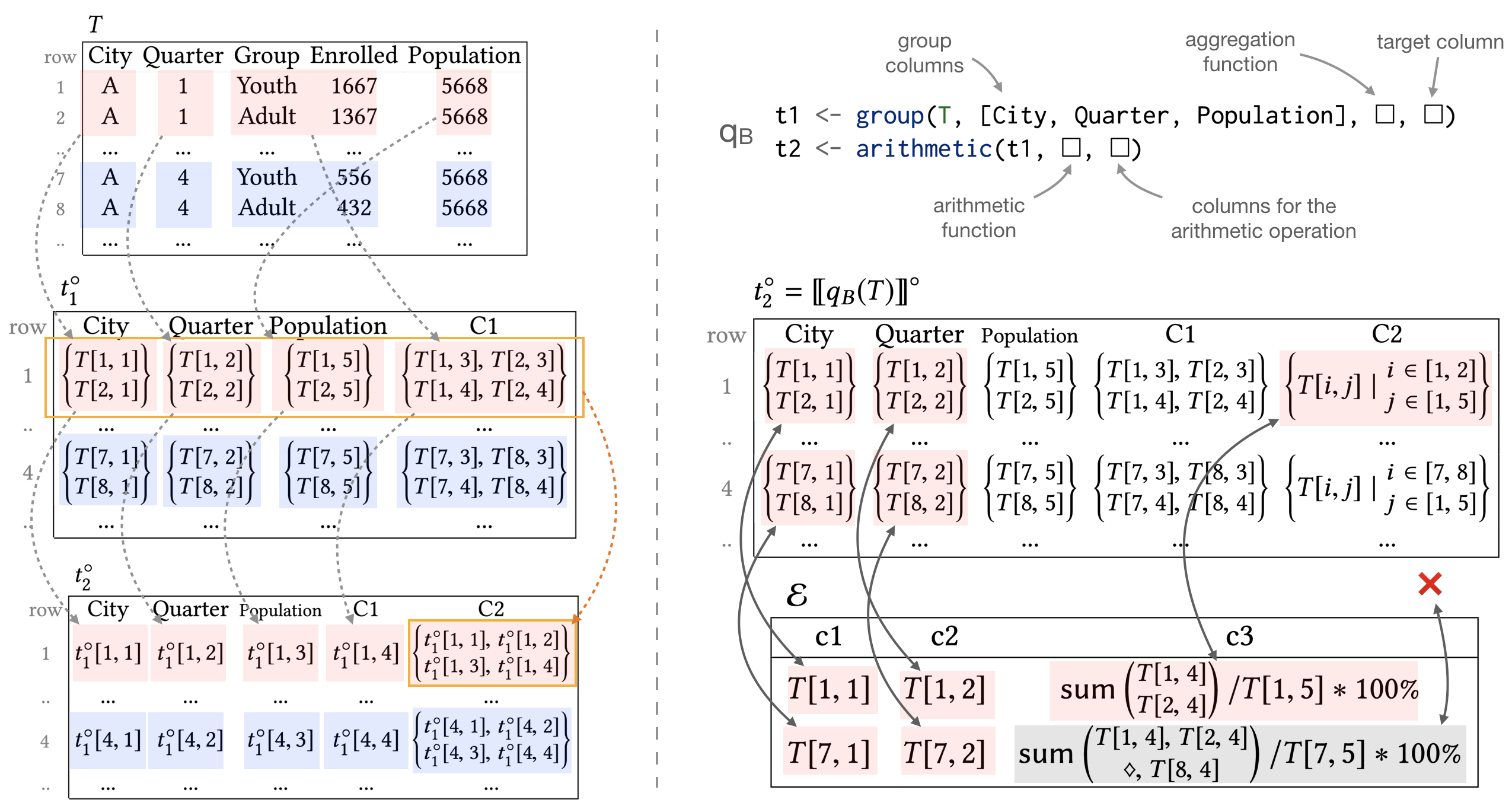}
\caption{Pruning the partial query $q_B$ using abstract provenance analysis. Right: $q_B$ can be pruned because its abstract output $t_2^\circ$ is inconsistent with the user demonstration $\mathcal{E}$. Left: how $t_2^\circ$ is derived from the input $T$ using abstract provenance analysis.}
\label{fig:overview-abstract-pruning}
\end{figure*}

\autoref{fig:overview-abstract-pruning} shows how our technique analyzes and prunes the incorrect partial query $q_B$ in \autoref{fig:overview-enumerative-search}. $q_B$ is a partial query with only group columns instantiated; other parameters are still ``$\square$''s (\autoref{fig:overview-abstract-pruning} right). Despite unknown parameters, we can analyze the data provenance abstractly (\autoref{fig:overview-abstract-pruning} left). 

First, for the subquery $q_{B1}$ {\mintinline[breaklines,fontsize=\footnotesize,escapeinside=||]{R}{t1 <- group(T, [City, Quarter, Population], |$\square$|, |$\square$|)}}: 
\begin{itemize}
	\item Because $T$ is grouped by \code{City}, \code{Quarter} and \code{Population}, the first two rows in $T$ are grouped together. Thus, $t_1^\circ[1,1]$ (the cell at row 1 column 1 from the output of $q_{B1}$) comes from $T[1,1], T[2,1]$, thus its provenance is $\left\{T[1,1], T[2,1]\right\}$; the provenances for $t_1^\circ[1,2]$ and $t_1^\circ[1,3]$ are derived similarly (shown by gray arrows). 
	\item Because the aggregation function and column are unknown, we \emph{over-approximate} the provenance of cells in $\mathsf{C1}$ (the new column generated by aggregation). For example, for $t_1^\circ[1,4]$, because the aggregation function can use any columns outside of the grouping columns, its abstract provenance is $S=\{T[1,3],\allowbreak T[2,3],\allowbreak T[1,4],\allowbreak T[2,4]\}$. This shows only values from the set $S$ will flow from $T$ to $t_1[1,4]$ regardless of how $q_{B1}$ is instantiated.
\end{itemize}
The analysis generates $t_1^\circ$, the abstract output of $q_{B1}$. Then, for the subquery $q_{B2}$ \colorbox{black!7}{\mintinline[fontsize=\footnotesize,escapeinside=||]{R}{t2 <- arithmetic(t1, |$\square$|, |$\square$|)}}:
\begin{itemize}
	\item Because $\mathsf{arithmetic}$ does not modify existing columns, columns 1,2,3,4 of $t_2^\circ$ are directly inherited from columns 1,2,3,4 of $t_1^\circ$ (e.g., the provenance of $t_2^\circ[1,1]$ is $\{t_1^\circ[1,1]\}$).
	\item Due to the unknown arithmetic function and its parameters, we need to over-approximate the provenance for cells in the new column. For $t_2^\circ[1,5]$, because any cells from row $1$ of $t_1^\circ$ can potentially be used in the arithmetic computation, its provenance is $\{t_1^\circ[1,1],\allowbreak t_1^\circ[1,2],\allowbreak t_1^\circ[1,3],\allowbreak t_1^\circ[1,4]\}$ (shown by the orange box in \autoref{fig:overview-abstract-pruning}).
\end{itemize}
This analysis generates the abstract output $t^\circ_2$. By combining the provenance information in $t_1^\circ$ and $t_2^\circ$, the final abstract output of $q_B$ is shown as $t_2^\circ$ in \autoref{fig:overview-abstract-pruning} right.

Given $t_2^\circ$, we check its consistency with the user demonstration $\mathcal{E}$: can we extract a subtable $t^\circ$ of $t_2^\circ$ such that there exists a one-to-one mapping between cells in $\mathcal{E}$ and $t^\circ$? This is a necessary condition $q_B$ need to satisfy to achieve the consistency goal from \autoref{subsec:overview-userspec}.

However, we cannot find such a consistent mapping between them, because the cell $\mathcal{E}[2,3]$ (colored grey) has no matching cell in $t_2^\circ$. The cell $\mathcal{E}[2,3]$ requires using at least values $\{T[1,4],\allowbreak T[2,4], \allowbreak T[8,4]\}$, but there is no cell in $t_2^\circ$ that contains these values, meaning that there is no way for these values in the input $T$ to flow into the same cell in the output $t_2^\circ$. Thus, no matter how $q_B$ is instantiated, it cannot realize our goal, and we can prune $q_B$. The same reasoning can prune many other infeasible queries in \autoref{fig:overview-enumerative-search}: the synthesizer visited only 1,453 (partial and concrete) programs before finding the solution $q_E$ within only 6 seconds.

Note that this abstract query $q_B$ cannot be pruned if we use value-based abstract semantics as in prior work~\cite{DBLP:conf/pldi/WangCB17,DBLP:journals/pacmpl/WangFBCD20} \emph{even if the user example contains full output and without incomplete expressions}: as shown in the alternative table $t_2^v$ below, we lack information about aggregation/arithmetic functions and their parameters, hence we cannot derive a concrete value to approximate the values in columns \code{C1} and \code{C2} (thus represented as {\color{gray} unknown}). These unknown values can match any value in $\mathcal{E}$, even with a full output example, the query $q_B$ cannot be pruned.

\begin{center}
{
\footnotesize
\setlength\tabcolsep{2pt}
\begin{tabular}{c|ccccc|}
\multicolumn{1}{l}{} & \multicolumn{5}{l}{$t_2^v$ (derived from $q_B(T)$ using value abstraction)\vspace{2pt}}\\
\cline{2-6}
{\color{gray}\footnotesize row}  & City	& Quarter &  Population & C1 & C2\\\cline{2-6}
\rowcolor{white!7} \cellcolor{white}{\color{gray}\footnotesize 1} & A		&   1    & 5668 & {\color{gray} unknown} & {\color{gray} unknown}\\
{\color{gray}\footnotesize ..}  & ...     & ...  & ...    & ...      & ... \\
\rowcolor{white!7} \cellcolor{white}{\color{gray}\footnotesize 4} & A    	&   4      & 5668 & {\color{gray} unknown} & {\color{gray} unknown}\\
{\color{gray}\footnotesize ..}  & ...     & ...  & ...    & ...      & ...\\\cline{2-6}
\end{tabular}
}\vspace{3pt}
\end{center}

%\subsection{Summary} 

%Here, we compare computation demonstration and I/O examples from the algorithm design perspective: (1) computation demonstration can be considered as a \emph{weaker specification} because it allows partial output and incomplete expressions, which makes type/value-based reasoning ineffective; (2) the consistency checking process is more expensive in computation demonstration (due to the overhead of provenance tracking evaluation and term equivalence checking), it requires stronger pruning to maintain efficiency; (3) however, because computation demonstration preserves finer-grained provenance information and maintains cell-level computation structure, it can also be considered a \emph{stronger specification} that enables new pruning opportunities.

%These properties drive our abstraction design: \tool leverages fine-grained abstract provenance and structure information to reason about abstract programs, and it discovers new pruning opportunities to prune infeasible programs for higher synthesis efficiency.

In summary, the user solves the analytical task using a computation demonstration. 
The benefit of our approach is that complex queries can be easily demonstrated in individual groups:
%permit simple demonstrations: 
the complexity of an analytical query often lies in how to join and partition the table properly, but in the demonstration, the user only needs to interact with values from \emph{one group}, thus avoiding the need to consider complex joining/grouping/partitioning behind the scenes.
\tool makes this new synthesis problem tractable by leveraging abstract provenance analysis for dramatic pruning.

We next formally present the synthesis specification (\autoref{sec:problem}), the abstraction-based pruning algorithm (\autoref{sec:algorithm}) and demonstrate the practical use of \tool through experiments on real world analytical SQL tasks (\autoref{sec:experiments}).

%% file: sections/problem.tex
\section{The Synthesis Problem}
\label{sec:problem}

In this section, we formally describe analytical SQL and the {program-by-computation-demonstration} problem. 

In the following, we use $q$ for queries, $T$ and $t$ for tables, and $c$ for table columns. We use the bar notation $\bar{x}$ to denote a list of $x$ (e.g., $\bar{T}$ stands for a list of tables $T_1,\dots, T_n$ and $\bar{c}$ for columns $c_1, \dots, c_n$), and $\bar{f}(\bar{x})$ for a list of function applications $[f_1(x_1),\dots,f_n(x_n)]$.

\subsection{Analytical SQL}

\paragraph{Table} A table is an ordered bag of tuples, and each tuple is a list of values (strings or numbers). A table $T$ with $n$ columns and $m$ rows is represented as $\{[v_{11},..., v_{1n}],..., [v_{m1},..., v_{mn}]\}$.

% \[
% \begin{array}{rcll}
%     T  & \leftarrow & \{\mathit{r}_1, \dots, \mathit{r}_n\} & \text{(Table)} \\
%     \mathit{r}  & \leftarrow & [{v}_1, \dots, {v}_m] & \text{(Row)} \\
% \end{array}
% \]

The choice of using ordered bag is to support sorting in intermediate computation steps, which is essential for order-dependent aggregation functions like $\mathsf{rank}$ and $\mathsf{cumsum}$. The ordered bag representation also permits the use of set operators as in normal bags, and allows us to refer to table rows based on their index (like lists). For simplicity, we omit table column names in our formulation, and table columns are referred based on column indexes. We use $T[i,j]$ to refer to the cell at row $i$ and column $j$, $T[i,*]$ for the $i$-th row in $T$, and $T[\bar{c}]$ for the result of projecting $T$ on columns $\bar{c}$. We also use $T_1\subseteq T_2$ to denote table containment, and $T_1\times T_2$ for table crossproduct. Two tables are equivalent ($T_1=T_2$) if they contain each other regardless of orders ($T_1\subseteq T_2 \land T_2\subseteq T_1$) (because row order only matters when used with order-dependent aggregation functions).

\paragraph{Analytical SQL} \autoref{fig:sql-grammar} defines the analytical SQL language $\mathcal{L}_\mathit{SQL}$ used in our paper. A query
$q$ is formed by compositions of basic constructors $\mathsf{proj}$ (projection), $\mathsf{filter}$, $\mathsf{join}$, $\mathsf{group}$, $\mathsf{partition}$, $\mathsf{sort}$ (sort rows), $\mathsf{aggregate}$, and $\mathsf{arithmetic}$ constructors. Comparing to prior work~\cite{DBLP:conf/pldi/WangCB17}, the key extensions are portioning, window functions ($\mathsf{cumsum}, \mathsf{rank}$) and arithmetic functions. Note that aggregation functions $\alpha$ can be used as analytical functions $\alpha'$ (but not otherwise).
%For instance, the query $\mathsf{proj}(q, [c_1, c_2])$ corresponds to ``\code{Select c1, c2 From q}''; the query $\mathsf{group}(q, c_1, \mathsf{sum}(c_2))$ corresponds to ``\code{Select c1, Sum(c2) From q Group By c1}'', and $\mathsf{partition}(q, c_1, \mathsf{cumsum}(c_2))$ corresponds to ``\code{Select c1, cumsum(c2) Over (Partition By c1) From q}''.
The semantics of analytical SQL follows the standard in modern databases. We use the notation $\eval{q(\bar{T})}$ to denote running $q$ on input tables $\bar{T}$, and the output is a table.

\begin{figure}[t]
\setlength\arraycolsep{3pt}
\[
\begin{array}{rcll}
    q  & \leftarrow & T ~|~  \mathsf{filter}(q, p) ~|~  \mathsf{join}(q_1, q_2) ~|~ \mathsf{proj}(q, \bar{c})) \\
        & | &  \mathsf{sort}(q, \bar{c}, \mathit{op}) ~|~ \mathsf{left\_join}(q_1, q_2, p) ~|~  \mathsf{arithmetic}(q, \gamma(\bar{c}))\\
        & | &  \mathsf{group}(q, \bar{c},  {\alpha}({c})) ~|~ \mathsf{partition}(q, {c},  {\alpha}'({c}))\\
    p   & \leftarrow & p_1~\mathsf{and}~p_2 ~|~\mathsf{true} ~|~ \mathsf{false} ~|~ c_1~\mathit{op}~c_2 \\
    \alpha & \leftarrow & \mathsf{sum} ~|~ \mathsf{avg} ~|~ \mathsf{max} ~|~ \mathsf{min} ~|~ \mathsf{count} \\
    \alpha' & \leftarrow & \alpha ~|~ \mathsf{dense\_rank} ~|~ \mathsf{rank} ~|~ \mathsf{cumsum}\\
    \mathit{op} & \leftarrow & < ~|~ \le ~|~ == ~|~ > ~|~ \ge\\
\end{array}
\]
\caption{The analytical SQL language, where $c$ denotes a column index, $\gamma$ refers to an arithmetic function.}
\label{fig:sql-grammar}
\end{figure}

\begin{figure}[t]
    \begin{smpage}{0.45\linewidth}
    \[
    \begin{array}{rcll}
        T^\star  & \leftarrow & [{r}^\star, \dots, {r}_n^\star]  \\
        {r}^\star  & \leftarrow & [{e}^\star_1, \dots, {e}^\star_m]  \\
        {e}^\star  & \leftarrow & \mathit{const} \\
            & | & T_k[i,j]\\
            & | & f({e}^\star_1, \dots, {e}^\star_l) \\
            & | & \mathsf{group}\{e^\star_1, \dots, e^\star_l\} \\
    \end{array}
    \]
    \end{smpage}\qquad
    \begin{smpage}{0.45\linewidth}
    \[
    \begin{array}{rcll}
        \mathcal{E}  & \leftarrow & [r_1, \dots, r_n]  \\
        r  & \leftarrow & [e_1, \dots, e_m]  \\
        e  & \leftarrow & \mathit{const} \\
            & | & T_k[i,j] \\
            & | & f(e_1, \dots, e_l) \\
            & | & f^\diamondsuit(e_1, \dots, e_l) \\
    \end{array}
    \]
    \end{smpage}
    \caption{Definitions of the provenance-embedded table $T^\star$ and the user demonstration $\mathcal{E}$. Metavariable $f$ refers to an aggregator or an arithmetic function, $T_k[i,j]$ denotes the reference to cell $i,j$ from the input table $T_k$, and ``$f^\diamondsuit$'' stands for a function $f$ with some parameters omitted by the user in their demonstration.}
    \label{fig:provenance-table-and-user-demo}
    \end{figure}
\begin{figure}[t]
\centering
\small
%\todo{write the evaluation rules for group, partition, and arithmetic}

\[
\eval{T}^\star = 
    {\left\{\big[ ``T[i,c]"\mid c\in\mathsf{columns}(T^\star)\big] \bigg|~ i\in [1,\mathsf{rowNum}(T^\star)] \right\}}
\]

\[
\begin{array}{ll}
\eval{\mathsf{filter}(q, p)(\bar{T})}^\star = ~\mathit{let}~T^\star=\eval{q(\bar{T})}^\star~\mathit{in}\\
    \qquad {\left\{ T^\star[i,*] \bigg|~ i\in [1,\mathsf{rowNum}(T^\star)],~ \eval{p(T^\star[i,*])} = \mathsf{true} \right\}}
\end{array}
\]

\[
\begin{array}{c}
\eval{\mathsf{proj}(q,\bar{c})(\bar{T})}^\star = \eval{q(\bar{T})}^\star[\bar{c}]\vspace{4pt}\\
\eval{\mathsf{join}(q_1, q_2)(\bar{T})}^\star = \eval{q_1(\bar{T})}^\star\times\eval{q_2(\bar{T})}^\star
\end{array}
\]

\[
\begin{array}{ll}
\eval{\mathsf{left\_join}(q_1, q_2, p)(\bar{T})}^\star = ~\mathit{let}~\left\{\begin{array}{ll}T_1^\star=\eval{q_1(\bar{T})}^\star\\ T_2^\star=\eval{q_2(\bar{T})}^\star\end{array}\right.~\mathit{in}\\
~{\left\{ T_1^\star[i,*] \concat T_2^\star[j,*] \bigg|~ \footnotesize\begin{array}{l}i\in [1,\mathsf{rowNum}(T_1^\star)],\\j\in [1,\mathsf{rowNum}(T_2^\star)]\\\eval{p(T_1^\star[i,*] \concat T_2^\star[j,*])}=\mathsf{true}\end{array} \right\}}\\
~\cup~{\left\{ T_1^\star[i,*] \concat \bar\emptyset~ \bigg|~ \footnotesize\begin{array}{l}i\in [1,\mathsf{rowNum}(T_1^\star)],~j\in [1,\mathsf{rowNum}(T_2^\star)]\\\eval{p(T_1^\star[i,*] \concat T_2^\star[j,*])}=\mathsf{false}\end{array} \right\}}
\end{array}
\]

\[
\begin{array}{ll}
\eval{\mathsf{sort}(q, \bar{c}, \mathit{op})(\bar{T})}^\star = ~\mathit{let}~T^\star=\eval{q(\bar{T})}^\star~\mathit{in}~\\
\qquad \left\{T^\star[k,*] ~|~ T^\star[k,\bar{c}]~\mathit{op}~ T^\star[k+1,\bar{c}], k\in [1,\mathsf{rowNum}(T^\star)] \right\}
\end{array}
\]

\[
\begin{array}{ll} 
\eval{\mathsf{group}(q, \bar{c}, \alpha(c_t))(\bar{T})}^\star = \\
    \qquad \mathit{let}~T^\star=\eval{q(\bar{T})}^\star,~G=\mathsf{extractGroups}(\eval{T^\star[\bar{c}]})~\mathit{in}\vspace{2pt}\\
    \qquad \qquad {\left\{\begin{array}{l} \big[\mathsf{group}\left\{ T^\star[k,c]\mid k\in g\right\} ~\big|~ c\in \bar{c}\big] \\\concat \big[\alpha(T^\star[k,c_{t}] \mid k\in g)\big]\end{array} \bigg|~ g\in G \right\}}
\end{array}
\]

\[
{\begin{array}{ll} 
    \eval{\mathsf{partition}(q, \bar{c}, \alpha'({c}_t))(\bar{T})}^\star = \\
\quad\mathit{let}~T^\star=\eval{q(\bar{T})}^\star,~G=\mathsf{extractGroups}(\eval{T^\star[\bar{c}]})~\mathit{in}\vspace{2pt}\\
    \qquad{\footnotesize\left\{\begin{array}{l} \big[T^\star[i,j]\mid j\in[1,\mathsf{colNum}(T^\star)]\big] \\\concat \left[\alpha'\left(T^\star[k,c_{t}] \;\middle|\; k,i\in g\in G \right) \right]\end{array} ~\middle|~ i\in [1,\mathsf{rowNum}(T^\star)] \right\}}
\end{array}}
\]

%{\tiny\begin{array}{l}g\ni i, \\g\in G,\\k\in g\end{array}}

\[
{\begin{array}{ll} 
    \eval{\mathsf{arithmetic}(q, \gamma, \bar{c})(\bar{T})}^\star = \mathit{let}~T^\star=\eval{q(\bar{T})}^\star~\mathit{in}\vspace{2pt}\\
    \quad\quad{\left\{\begin{array}{l}\big[T^\star[i,j]\mid j\in[1,\mathsf{colNum}(T^\star)]\big]\\ \qquad\concat \left[\gamma(T[i,c]\mid c\in\bar{c})\right] \end{array}\bigg|~ i\in [1,\mathsf{rowNum}(T^\star)] \right\}}
\end{array}}
\]

\[
\mathsf{extractGroup}(T) =\left\{S ~\bigg|~ \begin{array}{l}S\subseteq[1,\mathsf{rowNum}(T^\star)]\\ \forall i,j\in S, T[i,*]=T[j,*]\\\forall k\not\in S.\not\exists j\in S.T[k,*]= T[j,*]\end{array}\right\}
\]

\caption{Provenance tracking semantics of analytical SQL operators. The operator `$l_1\concat l_2$' concatenates two lists; the function $\mathsf{extractGroup}$ partitions the row indexes of table $T$ into disjoint equivalence sets (rows from the same set are equal, and rows from different sets are not equal).}
\label{fig:provenance-tracking-semantics}
\end{figure}

\paragraph{Provenance-Tracking Query Semantics} In addition to standard query evaluation rules, we introduce provenance-tracking semantics for analytical SQL. Under this semantics, each operator is a term rewriter that builds and transforms references to input table cells to keep track of the data provenance. Evaluating a query $q$ on inputs $\bar{T}$ under provenance-tracking semantics produces a \emph{provenance-embedded table} (denoted as $T^\star$). As shown in \autoref{fig:provenance-table-and-user-demo} left, each cell in $T^\star$ is an expression $e^\star$ that stores how the cell is derived from the inputs. A cell $e^\star$ is composed by $f$ (representing aggregation or arithmetic results), $\mathsf{group}\{\dots\}$ (results from the $\mathsf{group}$ operator), reference $T_k[i,j]$ and constants. A provenance embedded table $T^\star$ can be further evaluated into a table $T$ by evaluating the expressions in every table cell (denoted by $T=\eval{T^\star}$).
We use $\eval{q(\bar{T})}^\star$ to denote the provenance-tracking evaluation process. Rules for key SQL operators are shown in \autoref{fig:provenance-tracking-semantics}. For example, for the $\mathsf{partition}(q, \bar{c}, \alpha'({c}_t))$ operator, the rule first partitions the input $T^\star$ based on its partitioning columns $\bar{c}$ (using the auxiliary function $\mathsf{extractGroups}$). Then, a new aggregated value is generated for each row $i$: the rule finds the group $g$ that row $i$ belongs to it and constructs an expression $\alpha'(\dots)$ that includes all values in column $c_{t}$ from the group $g$. Also, in the evaluation process, \tool simplifies consecutive applications of aggregation functions like $f(f(a, b), c)$ into $f(a, b, c)$ for  $f\in\{\mathsf{sum}, \mathsf{max}, \mathsf{min}\}$. This let \tool better reason about semantically equivalent aggregations (we do not consider general term equivalence checking as we focus on analytical SQL; in future, our approach can also work with advanced equivalence checker~\cite{willsey2021egg} when applied to new domains).

\subsection{The Synthesis Task} 

We now define the specification language in which the user provides demonstrations and formulates the synthesis task. 

\paragraph{The User Specification} Besides input tables $\bar{T}$, the user provides a demonstration $\mathcal{E}$ (which is a table) consists of \emph{partial expressions} showing how each output cell is derived from inputs. As shown in \autoref{fig:provenance-table-and-user-demo}, each cell in $\mathcal{E}$ is an expression $e$, constructed from either a constant, a reference to the cell $i,j$ in input table $T_k$ ($T_k[i,j]$), an expression, or a partial expression with some omitted values. A function with omitted values is denoted as $f^\diamond(e_1,\dots,e_l)$, meaning that the user only provides values $e_1, ..., e_l$ and leaves out some others: the omitted values can appear at either beginning, middle or end of the function (e.g., in \autoref{fig:overview-user-demonstration}, omitted values are in the middle of the $\mathsf{sum}$ expression). Comparing to $T^\star$, cells in $\mathcal{E}$ are not expected to be constructed from $\mathsf{group}\{\dots\}$: the design rationale is that all cells in the same $\mathsf{group}$ have the same value; thus, the user only needs to use {any} one of them in the demonstration (as shown in \autoref{fig:overview-user-demonstration}). %For example, $T[1,1]$ and $T[2,1]$ in \autoref{eg:prov-tracking-eval} belongs to the same group when evaluated under $q_A$, because their values are both $1$, and the user only need to use one of them in the demonstration.

% \begin{center}
% \small
% \begin{tabular}{|cc|}
% \hline
% c1 & c2\\\hline
% $T[1,1]$ & ${\color{blue}\mathsf{sum}(}T[1,3],\square{\color{blue})}$ \\\hline
% \end{tabular}
% \end{center}

\paragraph{Consistency with the User Specification} To define the synthesis task, we first need to define what properties a provenance-embedded tables $T^\star$ needs to satisfy to be considered consistent with the user demonstration $\mathcal{E}$.

We first define cell-level consistency criteria: we say an expression $e^\star$ (from a provenance-embedded table) is consistent with an expression $e$ (from the user demonstration $\mathcal{E}$) if $e$ can be derived from $e^\star$ using rules from \autoref{fig:value-generalization} (denoted as $e\prec e^\star$). Concretely, $e^\star$ is consistent with $e$ if (1) $e^\star$ and $e$ are the same, (2) $e^\star$ is of form $\mathsf{group}\{\bar{e}^\star\}$ and a member $e^\star_i$ is consistent with $e$, or (3) $e^\star$ and $e$ are both expressions $f(\dots)$ and arguments in $e$ are subsumed by those in $e^\star$. In case (3), parameter order is not enforced if $f$ is commutative (e.g., aggregation function $\mathsf{sum}()$; arithmetic function $\lambda x,y.x*y$).

Table-level consistency rules are defined in \autoref{def:table-generalization}. Intuitively, a provenance-embedded table $T^\star$ is consistent with the user demonstration $\mathcal{E}$ if there exists a subtable $T^\star_1\subseteq T^\star$ satisfying the property that each cell $T^\star_1[i,j]$ generalizes the cell $\mathcal{E}[i,j]$ in the user demonstration.

\begin{definition}{\emph{(Provenance Consistency)}}\label{def:table-generalization}
Given a provenance-embedded table $T^\star$ and the user demonstration $\mathcal{E}$ with $m$ rows and $n$ columns, $T^\star$ is provenance-consistent with $\mathcal{E}$ if: $\exists T_1^\star \subseteq T^\star.~\forall\!\!\!{\scriptsize\begin{array}{ll}i\in[1,n]\\j\in[1,m]\end{array}}\!\!\!.~\mathcal{E}[i,j] \prec T_1^\star[i,j]$.
\end{definition}

\begin{figure}[t]
\centering
\footnotesize

\AXC{\vphantom{const}}
\UIC{$\mathit{const} \prec \mathit{const}$}
\DP
\quad
\AXC{\vphantom{const}}
\UIC{$T[i,j] \prec T[i,j]$}
\DP
\quad 
\AXC{$\exists e_i^\star \in \{\bar{e}^\star\}. e \prec e_i^\star$}
\UIC{$e \prec \mathsf{group}\{\bar{e}^\star\}$ }
\DP

\vspace{10pt}

{\scriptsize \AXC{{$\mathsf{isCommutative}(f)$}\quad{$\exists.~\{e^\star_{k_1}\dots e^\star_{k_m}\}\subseteq\{e^\star_1\dots e^\star_n\}.~\forall i\in[1,m].e_i \prec e^\star_{k_i}$}}
\UIC{$f^\diamondsuit(e_1,\dots, e_m) \prec f(e^\star_1,\dots, e^\star_n)$}
\DP}

\vspace{10pt}

{\scriptsize 
\AXC{\stackunder{$\neg\mathsf{isCommutative}(f)$}{$\forall i\in[1,m].e_i \prec e^\star_{i}$}}
\UIC{$f^\diamondsuit(e_1,\dots, e_m) \prec f(e^\star_1,\dots, e^\star_n)$}
\DP
\quad 
\AXC{$f^{\diamondsuit}(e_1,\dots, e_n) \prec f(e^\star_1,\dots, e^\star_n)$}
\UIC{$f(e_1,\dots, e_n) \prec f(e^\star_1,\dots, e^\star_n)$}
\DP}

\caption{Rules that determine whether an expression $e^\star$ from a provenance-embedded table generalizes an expression $e$ from the user demonstration ($e \prec e^\star$).}
\label{fig:value-generalization}
\end{figure}

Finally, we define the task: synthesize queries whose outputs are provenance consistent with the user demonstration. 

\begin{definition}{\emph{(The Synthesis Task)}}\label{def:synthesis-task}
Given input tables $\bar{T}$ and user demonstration $\mathcal{E}$, the synthesis task is to find analytical SQL queries, such that each $q$ satisfies $\mathcal{E}\prec\eval{q(\bar{T})}^\star$.
\end{definition}

\noindent \emph{Remarks.} Since the user demonstration is an incomplete specification of the actual task, it is inherently ambiguous, and there could be multiple queries consistent with the user specification. As we will present more in detail in \autoref{sec:experiment-setup}, our algorithm addresses this problem by synthesizing a set of queries consistent with the demonstration and present them to the user; it can work with existing program disambiguation framework to interactively resolve ambiguity (which is not the focus of our paper).

%% file: sections/algorithm.tex
%!TEX root=../main.tex

\section{The Synthesis Algorithm}
\label{sec:algorithm}

In this section, we introduce our synthesis algorithm: given input tables $\bar{T}$ and the user demonstration $\mathcal{E}$, our algorithm aims to find a set of queries consistent with $\mathcal{E}$.

%\subsection{The Top Level Algorithm: Enumerative Search}

\begin{algorithm}[t]
  \begin{algorithmic}[1]
\Procedure{Synthesize}{$\bar{T}, \mathcal{E}, \mathit{depth}, \mathit{N}$}
  \State {\rm \bf input:} input tables $\bar{T}$, user demonstration $\mathcal{E}$, search depth $\mathit{depth}$, query number limit $\mathit{N}$.
  \State {\rm \bf output:} a set of queries consistent with $\bar{T}$ and $\mathcal{E}$
  \bigskip

% \State $W_0 \gets \{T_1, \dots, T_n\}$;
% \State $k\gets 0$;
% \While{$k < \mathit{depth}$}
%   \For{$\mathit{\gamma}\in{\small\left\{\begin{array}{ll}\mathsf{group\_by}, \mathsf{filter}, \mathsf{partition\_by},\\ \mathsf{proj},\mathsf{aggregate}, \mathsf{arithmetic}, \mathsf{sort} \end{array}\right\}}$}
%     \State $W_k \leftarrow \{\gamma({q}, \bar\square) ~|~ {q}\in W_{k-1}\}$;
%   \EndFor
%   \For{$\mathit{\gamma}\in\{\mathsf{join}, \mathsf{left\_join}\}$}
%   \State $W_k \leftarrow W_k \cup \left\{\gamma({q}_1, {q}_2, \square) ~|~ {\scriptsize\begin{array}{ll}{q}_1 \in W_i, {q}_2\in W_j\\i + j = k-1\end{array}}\right\}$;
%   \EndFor
% \EndWhile
% \State $W\leftarrow \bigcup_{i\in[0, \mathit{depth}]} W_i$;

\State $W\leftarrow \mathsf{constructSkeletons}(\bar{T}, \mathit{depth})$;
\State $R\leftarrow \emptyset$;
  \While{$\neg W.\mathsf{isEmpty}()$}
    \State ${q} \gets W.\mathsf{next}()$;
    \If{$\mathsf{IsConcrete}({q})$ }
    \If{$\mathcal{E}\prec\eval{{q}(\bar{T})}^\star$}   
      \State $R\leftarrow R \cup \{{q}\}$;
      \If{$|R| \ge \mathit{N}$} \Return $R$
      \EndIf
    \Else \ {\bf continue};\EndIf
    \EndIf
    \State $\phi \gets \mathsf{AbstractReasoning}({q}, \bar{T}, \mathcal{E})$;
    \If{$\mathsf{UNSAT}(\phi)$} $\mathbf{continue}$; \EndIf
    \State  $\square_i \gets \mathsf{chooseNextHole}({q})$;
    \State $D \gets \mathsf{inferDomain}(\square_i, {q}, \bar{T})$;
  \State $W \gets W \cup \left\{ [\square_i \mapsto v]{q} \ \big{|} \ v \in D \right\}$
   % \For{$v \in \mathsf{dom}(\square_k)$}
   % \State $\worklist \gets \worklist \cup \partialprog[\square_k \mapsto v]$
   % \EndFor
  \EndWhile
\Return $R$
\EndProcedure
  \end{algorithmic}
  \caption{Top-level analytical SQL synthesis algorithm.}
  \label{alg:top-level}
\end{algorithm}

\autoref{alg:top-level} shows our top-level synthesis algorithm: an abstraction-based enumerative search algorithm that iteratively enumerates and prunes partial programs until up-to-N programs consistent with the user specification are found. %The \textsc{Synthesize} procedure contains two main components: (1) a coarse search component that enumerates query skeletons -- queries with no instantiated parameters, and (2) a fine search phase that fills parameters of feasible query skeletons until satisfying queries are found. 

\textsc{Synthesize} starts out by constructing query skeletons from input tables $\bar{T}$ up to size $\mathit{depth}$ (line 4). A query skeleton is composed from top-level query operators in \autoref{fig:sql-grammar} but with all arguments unfilled (represented as holes ``$\square$'').  %the algorithm constructs query skeletons of size $k$ by applying unary (line 7-8) or binary (9-10) constructors on previously generated skeletons. This process generates the set of all query skeletons $W$ up to size $\mathit{depth}$ (an algorithm parameter).
Then, the algorithm takes one partial query ${q}$ out of the work list $W$ at a time. If the query ${q}$ is already concrete (i.e., no hole left to be instantiated), the algorithm checks if it is consistent with the user demonstration $\mathcal{E}$ and add it to the result set $R$ if so (line 8-12). If ${q}$ is partial, the algorithm conducts an abstract analysis to check if $q$ can potentially realize the user demonstration $\mathcal{E}$. If the check fails, the partial query is pruned (line 14). Otherwise, the synthesizer expands the search space by (1) choosing the next hole $\square_i$ in ${q}$ to be instantiated, (2) inferring the domain of the hole $\square_i$, and (3) replacing the hole $\square_i$ with each candidate value $v$ and add it to $W$ for further search (lines 15-17).

While the \textsc{Synthesize} procedure is a sound and complete way to explore the query space within the given depth, its efficiency relies on whether the abstract reasoning subroutine $\mathsf{AbstractReasoning}$ can effectively prune infeasible partial programs early to avoid search explosion. 
%\subsection{Pruning with Abstract Provenance Analysis}
As discussed in \autoref{sec:intro}, our key insight is to abstractly reason about cell-level provenance of the output table to check whether required values can flow from the input data to it in a way consistent with the user specification $\mathcal{E}$.

\begin{figure}[t]
\centering
\footnotesize
%\todo{write the evaluation rules for group, partition, and arithmetic}
\[
\eval{T_k}^\circ = 
    {\left\{\big[ \left\{``T_k[i,c]"\right\}\mid c\in \mathsf{columns}(T_k)\big] ~\middle|~ i\in [1,\mathsf{rowNum}(T_k)] \right\}}
\]

\[
\begin{array}{c}
\eval{\mathsf{filter}(q, \square)(\bar{T})}^\circ = \eval{q(\bar{T})}^\circ \qquad \eval{\mathsf{proj}(q, \square)(\bar{T})}^\circ = \eval{q(\bar{T})}^\circ \vspace{4pt}\\
\eval{\mathsf{join}(q_1, q_2)(\bar{T})}^\circ = \eval{q_1(\bar{T})}^\circ \times \eval{q_2(\bar{T})}^\circ
\end{array}
\]

\[
\begin{array}{ll} 
\eval{\mathsf{left\_join}(q_1, q_2, \square)(\bar{T})}^\circ = \mathit{let}~T_1^\circ=\eval{q_1(\bar{T})},~T_2^\circ=\eval{q_2(\bar{T})}^\circ~\mathit{in}\vspace{2pt}\\
\qquad (T_1^\circ\times T_2^\circ) \cup \{T_1^\circ[i,*]+\bar{\emptyset}~|~i\in[1,\mathsf{rowNum}(T_1^\circ)]\}
\end{array}
\]

\[
\begin{array}{ll}
\eval{\mathsf{sort}(q, \bar{c}, \mathit{op})(\bar{T})}^\circ = ~\mathit{let}~T^\circ=\eval{q(\bar{T})}^\circ~\mathit{in}~\\
\qquad \left\{T^\circ[k,*] ~|~ T^\circ[k,\bar{c}]~\mathit{op}~ T^\circ[k+1,\bar{c}], k\in [1,\mathsf{rowNum}(T^\circ)] \right\}~\\
\mathit{(when~T^\circ[\bar{c}]~is~concrete)}
\end{array}
\]

\vspace{6pt}
\dotfill\colorbox{black!4}{\small \emph{Weak abstraction (no instantiated parameters)}}\dotfill
\[
\begin{array}{ll} 
\eval{\mathsf{arithmetic}(q, \square, \square)(\bar{T})}^\circ =~\mathit{let}~T^\circ=\eval{q(\bar{T})}^\circ~\mathit{in}\vspace{2pt}\\
  \qquad {\left\{\begin{array}{l} \big[ T^\circ[i,c] ~\big|~ c\in \mathsf{columns}(T^\circ)\big] \\\concat \left[\bigcup \big\{T^\circ[i,c]\mid c\in \mathsf{columns}(T^\circ)\big\}\right]\end{array} \bigg|~ i\in[1,\mathsf{rowNum}(T^\circ)] \right\}}
\end{array}
\]

\[
\begin{array}{ll} 
\eval{\mathsf{partition}(q, \square, \square(\square))(\bar{T})}^\circ=~{\footnotesize \mathit{let}~\left\{{\arraycolsep=1.4pt\begin{array}{l}T^\circ=\eval{q(\bar{T})}^\circ\\N=\mathsf{rowNum}(T^\circ)\end{array}}\right.~\mathit{in}}\vspace{2pt}\\
    \qquad {\left\{\begin{array}{l} \big[\left\{ T^\circ[i,c]\right\} ~\big|~ c\in \mathsf{columns}(T^\circ)\big] \\\concat \left[\bigcup\big\{T^\circ[k,c]\mid k\in [1,N], c\in \mathsf{columns}(T^\circ)\big\}\right]\end{array} \bigg|~ i\in[1,N] \right\}}
\end{array}
\]

% \[
% \eval{\mathsf{group}(q, \square, \square(\square))(\bar{T})}^\circ=\eval{\mathsf{partition}(q, \square, \square(\square))(\bar{T})}^\circ
% \]

\[
\begin{array}{ll} 
\eval{\mathsf{group}(q, \square, \square(\square))(\bar{T})}^\circ=~{\footnotesize \mathit{let}~T^\circ=\eval{q(\bar{T})}^\circ,~N=\mathsf{rowNum}(T^\circ)~\mathit{in}}\vspace{2pt}\\
    \qquad {\left\{\begin{array}{l} \big[ \bigcup\{T^\circ[k,c] \mid k\in [1,N]\} ~\big|~ c\in \mathsf{columns}(T^\circ)\big] \\\concat \left[\bigcup\big\{T^\circ[k,c]\mid k\in [1,N], c\in \mathsf{columns}(T^\circ)\big\}\right]\end{array} \bigg|~ i\in[1,N] \right\}}
\end{array}
\]

\vspace{6pt}
\dotfill\colorbox{black!4}{\small \emph{Medium-precision abstraction (some known parameters)}}\dotfill
\[
\begin{array}{ll} 
  \eval{\mathsf{arithmetic}(q, \square, \bar{c})(\bar{T})}^\circ = \mathit{let}~T^\circ=\eval{q(\bar{T})}^\circ,~\mathit{in}\vspace{2pt}\\
    \qquad {\left\{ \begin{array}{l} \big[ T^\circ[i,c] ~\big|~ c\in \mathsf{columns}(T^\circ)\big] \\\concat \left[\bigcup\big\{T^\circ[i,c]\mid c\in \bar{c}\big\}\right]\end{array} \bigg|~ i\in[1,\mathsf{rowNum}(T^\circ)\} \right]}
\end{array}
\]

% {\footnotesize
% \[
% \eval{\mathsf{partition}(q, \bar{c}, \square(\square))(\bar{T})}^\circ = \left\{\begin{array}{ll}
% \begin{array}{ll} 
%     \mathit{let}~T^\circ=\eval{q(\bar{T})}^\circ,~N=\mathsf{rowNum}(T^\circ)~\mathit{in}\vspace{2pt}\\
%     \quad {\left[\begin{array}{l} \big[ T^\circ[i,c] ~\big|~ c\in \mathsf{columns}(T^\circ)\big] \\\concat \left[\bigcup\big\{T^\circ[k,c']\mid k\in [1,N], c'\not\in\bar{c}\big\}\right]\end{array} \bigg|~ i\in[1,N] \right]}
% \end{array} &(T^\circ[\bar{c}]~\textit{is abstract})\vspace{10pt}\\
% \begin{array}{ll} 
%     \mathit{let}~\left\{\begin{array}{l}T^\circ=\eval{q(\bar{T})}^\circ,N=\mathsf{rowNum}(T^\circ)\\G=\mathsf{extractGroups}(\eval{T^\circ[\bar{c}]})\end{array}\right.~\mathit{in}\vspace{4pt}\\
%     \quad {\left[\begin{array}{l} \big[ T^\circ[i,c] ~\big|~ c\in \mathsf{columns}(T^\circ)\big] \\\concat \left[\bigcup\left\{T^\circ[k,c']~\middle|~ {\tiny\begin{array}{l}g\ni i, \\g\in G,\\k\in g\end{array}}, c'\not\in\bar{c}\right\}\right]\end{array} \bigg|~ i\in[1,N] \right]}
% \end{array}&(T^\circ[\bar{c}]~\textit{is concrete})
% \end{array}
% \right.
% \]}

\[
\begin{array}{ll} 
  \eval{\mathsf{partition}(q, \bar{c}, \square(\square))(\bar{T})}^\circ =~\mathit{let}~\left\{{\arraycolsep=1.4pt\begin{array}{l}T^\circ=\eval{q(\bar{T})}^\circ\\N=\mathsf{rowNum}(T^\circ)\end{array}}\right.~\mathit{in}\vspace{2pt}\\
    \qquad {\left\{\begin{array}{l} \big[ T^\circ[i,c] ~\big|~ c\in \mathsf{columns}(T^\circ)\big] \\\concat \left[\bigcup\big\{T^\circ[k,c']\mid k\in [1,N], c'\not\in\bar{c}\big\}\right]\end{array} \bigg|~ i\in[1,N] \right\}}
\end{array} %&(T^\circ[\bar{c}]~\textit{is abstract})\vspace{10pt}
\]

\[
\begin{array}{ll} 
\eval{\mathsf{group}(q, \bar{c}, \square(\square))(\bar{T})}^\circ =~\mathit{let}~T^\circ=\eval{q(\bar{T})}^\circ,~N=\mathsf{rowNum}(T^\circ)~\mathit{in}\vspace{2pt}\\
    \qquad {\left\{\begin{array}{l} \big[ \bigcup\{T^\circ[k,c] \mid k\in [1,N]\} ~\big|~ c\in \bar{c}\big] \\\concat \left[\bigcup\big\{T^\circ[k,c']\mid k\in [1,N], c'\not\in\bar{c}\big\}\right]\end{array} \bigg|~ i\in[1,N] \right\}}
\end{array} %&(T^\circ[\bar{c}]~\textit{is abstract})\vspace{10pt}\\
\]

\vspace{6pt}
\dotfill\colorbox{black!4}{\small \emph{Strong abstraction ($T^\circ[\bar{c}]$ is concrete)}}\dotfill
\[
\begin{array}{ll} 
  \eval{\mathsf{partition}(q, \bar{c}, \square(\square))(\bar{T})}^\circ =~\mathit{let}~\left\{{\arraycolsep=1.4pt\begin{array}{l}T^\circ=\eval{q(\bar{T})}^\circ\\G=\mathsf{extractGroups}(\eval{T^\circ[\bar{c}]})\end{array}}\right.~\mathit{in}\vspace{4pt}\\
    \quad {\left\{\begin{array}{l} \big[ T^\circ[i,c] ~\big|~ c\in \mathsf{columns}(T^\circ)\big] \\\concat \left[\bigcup\left\{T^\circ[k,c']~\middle|~ {\tiny i,k\in g\in G}, c'\not\in\bar{c}\right\}\right]\end{array} \bigg|~ i\in[1,\mathsf{rowNum}(T^\circ)] \right\}}
\end{array}%&(T^\circ[\bar{c}]~\textit{is concrete})
\]

\[
\begin{array}{ll} 
\eval{\mathsf{group}(q, \bar{c}, \square(\square))(\bar{T})}^\circ =~\mathit{let}~\left\{{\arraycolsep=1.4pt\begin{array}{l}T^\circ=\eval{q(\bar{T})}^\circ\\G=\mathsf{extractGroups}(\eval{T^\circ[\bar{c}]})\end{array}}\right.~\mathit{in}\vspace{2pt}\\
    \qquad {\left\{\begin{array}{l} \big[\bigcup\left\{ T^\circ[k,c]\mid k\in g\right\} ~\big|~ c\in \bar{c}\big] \\\concat \left[\bigcup\big\{T^\circ[k,c']\mid k\in g, c'\not\in\bar{c}\big\}\right]\end{array} \bigg|~ g\in G \right\}}
\end{array}%&(T^\circ[\bar{c}]~\textit{is concrete})
\]
% {\footnotesize
% \[
% \eval{\mathsf{group}(q, \bar{c}, \square(\square))(\bar{T})}^\circ = \left\{\begin{array}{ll}
% \begin{array}{ll} 
%     \mathit{let}~T^\circ=\eval{q(\bar{T})}^\circ,~N=\mathsf{rowNum}(T^\circ)~\mathit{in}\vspace{2pt}\\
%     \quad {\left[\begin{array}{l} \big[ \bigcup\{T^\circ[k,c] \mid k\in [1,N]\} ~\big|~ c\in \bar{c}\big] \\\concat \left[\bigcup\big\{T^\circ[k,c']\mid k\in [1,N], c'\not\in\bar{c}\big\}\right]\end{array} \bigg|~ i\in[1,N] \right]}
% \end{array} &(T^\circ[\bar{c}]~\textit{is abstract})\vspace{10pt}\\
% \begin{array}{ll} 
%     \mathit{let}~T^\circ=\eval{q(\bar{T})}^\circ,~G=\mathsf{extractGroups}(\eval{T^\circ[\bar{c}]})~\mathit{in}\vspace{2pt}\\
%     \quad {\left[\begin{array}{l} \big[\bigcup\left\{ T^\circ[k,c]\mid k\in g\right\} ~\big|~ c\in \bar{c}\big] \\\concat \left[\bigcup\big\{T^\circ[k,c']\mid k\in g, c'\not\in\bar{c}\big\}\right]\end{array} \bigg|~ g\in G \right]}
% \end{array}&(T^\circ[\bar{c}]~\textit{is concrete})
% \end{array}
% \right.
% \]
% }

\caption{The abstract semantics for analytical SQL operators that over-approximates cell provenance. 
%The condition ``$T^\circ[\bar{c}]$'' checks whether the subset of the table contains sufficient information to derive concrete values, which can be used to decide grouping/partitioning information to obtain more precise analysis results. 
The analysis is more precise when more parameters are instantiated.}
\label{fig:abstract-provenance-semantics}
\end{figure}

\paragraph{The Abstract Semantics} Given a partial query $q$ and input tables $\bar{T}$, the abstract provenance analyzer returns an abstract output table $T^\circ$, where each cell $T^\circ[i,j]$ stores a set of references to input table cells that \emph{over-approximates} of all possible input cells that can flow to $T^\circ[i,j]$. \autoref{fig:abstract-provenance-semantics} shows the abstract semantics of key  operators.
%\footnote{Tor conciseness, we only formulate rules that considering one aggregation function - target column pair; the analysis of a query with with $n$ aggregation function - target column pairs can be achieved by applying the analysis rule $n$ times.}: 

In the base case, given an input table $T_k$, each cell in the abstract output symbolically stores the table id and its row/column indexes ($``T_k[i,j]"$). Because $\mathsf{filter}$, $\mathsf{proj}$, $\mathsf{join}$, $\mathsf{left\_join}$ and $\mathsf{sort}$ do not derive new cells, they simply propagate abstract analysis results of their subqueries to top-level operators. For core analytical operators $\mathsf{group}$, $\mathsf{partition}$ and $\mathsf{arithmetic}$, different levels of abstraction are designed based on concreteness of operator to accommodate analysis for queries in different stages in the search process --- the abstraction is stronger when more parameters are instantiated. For example, the abstraction of $\mathsf{partition}$ is defined as follows: 
\begin{itemize}
  \item \emph{Weak abstraction:} When the query is fully abstract (i.e., $\mathsf{partition}(q,\square,\square(\square))$), the analyzer can only confirm that existing columns in $T^\circ$ will be preserved, and every value in the newly generated column can use \emph{any} cell from $T^\circ$. While it is weak, it can propagate abstraction from $T^\circ$.
  \item \emph{Medium-precision abstraction:} When the partition columns $\bar{c}$ are known ($\mathsf{partition}(q,\bar{c},\square(\square))$) but $T^\circ[\bar{c}]$ is abstract, the analyzer can narrow down aggregation columns (to those outside $\bar{c}$). Thus, the new value in row $i$ can refer to any values from any rows in $T^\circ$ that is not in columns $\bar{c}$. %But since concrete values are still unavailable in $T^\circ[\bar{c}]$, the analyzer cannot decide how the table is partitioned,
  \item \emph{Strong abstraction:} On top of the last case, the subquery $T^\circ[\bar{c}]$ is now concrete, and this lets the analyzer know how to partition the table $T^\circ$. This refines the provenance of the newly generated value at row $i$: it can only include values from rows that are in the group as row $i$ and are outside of columns $\bar{c}$. This is a strong abstraction.
\end{itemize}
The abstract semantics for $\mathsf{arithmetic}$ and $\mathsf{group}$ follows the same principle: when the analyzer encounters a concrete subquery $q$ (free of holes), the analyzer will evaluate $q$ using provenance-tracking semantics and pass the concrete output for further abstract reasoning to achieve stronger analysis; otherwise, the rule propagates and builds abstract provenance on top of abstract analysis results from its subqueries. 

When an abstract query $q$ cannot pattern match any of the abstract analysis rules, the analysis process stops and returns $\mathsf{SAT}$ to the main synthesis algorithm (line 13 in \autoref{alg:top-level}): in such case $q$ is too abstract to derive a strong analysis results for pruning, and it is the main synthesis loop's responsibility to further instantiate the query.

\paragraph{Provenance Consistency} Given an abstract query $q$, we use the following criteria to check whether the query is consistent with the input $\bar{T}$ and the user demonstration $\mathcal{E}$:

\begin{definition}{\emph{(Abstract Provenance Consistency)}}\label{def:abstract-consistency}
Given a table $T^\circ$ (generated by the abstract analyzer) and the user demonstration $\mathcal{E}$ with $m$ rows and $n$ columns, $T^\circ$ is consistent with $\mathcal{E}$ if the following predicate is true (denoted as  $\mathcal{E} \lhd T^\circ$): 
$$\exists r_1\dots r_m,c_1\dots c_n.~\forall\!\!\!{\scriptsize\begin{array}{ll}i\in[1,n]\\j\in[1,m]\end{array}}\!\!\!.~\mathsf{ref}(\mathcal{E}[i,j]) \subseteq T^\circ[r_i, c_j]$$ 
\end{definition}

\noindent Here, the $\mathsf{ref}$ function extracts all table references in a cell $e$ ($\mathsf{ref}$ also works for $e^\star$ defined in \autoref{fig:provenance-table-and-user-demo}):
{\small\[
\begin{array}{ll}
\mathsf{ref}(T_k[i,j]) = \{T_k[i,j]\} & \mathsf{ref}(f^\diamondsuit(e_1, \dots, e_l)) = \bigcup_{i\in[1,l]}\{\mathsf{ref}(e_i)\}\\
\mathsf{ref}(\mathit{const}) = \emptyset & \mathsf{ref}(f(e_1, \dots, e_l)) = \bigcup_{i\in[1,l]}\{\mathsf{ref}(e_i)\}\\
\multicolumn{2}{l}{\mathsf{ref}(\mathsf{group}\{e_1, \dots, e_l\}) = \bigcup_{i\in[1,l]}\{\mathsf{ref}(e_i)\}}
\end{array}
\]}

\noindent We next show how the checking criteria can be used to prune infeasible partial programs.

\begin{property}{\emph{(The relation between $\eval{}^\star$ and $\eval{}^\circ$)}}\label{prop:relation}
Given an abstract query $q$ and input tables $\bar{T}$, let $S_q$ be the set of all possible queries that can be instantiated from $q$, then:
\[\begin{array}{l}
  \forall q'\in S_q.~\exists r_1\dots r_m,c_1\dots c_n.~\forall i\in[1,m],j\in[1,n].\\
  \qquad\qquad\qquad~\mathsf{ref}(\eval{q'(\bar{T})}^\star[i,j]) \subseteq \eval{q(\bar{T})}^\circ[r_i,c_j]
\end{array}\]
\noindent where $m=\mathsf{rowNum}(\eval{q'(\bar{T})}^\star), n=\mathsf{colNum}(\eval{q'(\bar{T})}^\star)$.
\end{property}

\noindent This property is witnessed by the comparison between provenance tracking evaluation rules in \autoref{fig:provenance-tracking-semantics} and abstract evaluation rules in \autoref{fig:abstract-provenance-semantics}: for each row $i$ in $T^\star$, there exists a row $i'$ in $T^\circ$ such that each cell $T^\star[i,j]$ can map to a cell $T^\star[i',k]$ that contains a superset of input cells used by $T^\star[i,j]$. Note that $\eval{}^\circ$ can produce more columns and rows than $\eval{}^\star$ because it does not filter or project any newly generated column or rows.

\begin{property}{\emph{(Pruning Foundation)}}\label{prop:soundness}
Given an abstract query $q$ and input tables $\bar{T}$, let $S_q$ be the set of all possible queries that can be instantiated from $q$, then:
$$\mathcal{E} \not\triangleleft \eval{q(\bar{T})}^\circ \Longrightarrow \not\exists q'\in S_q.\mathcal{E}\prec \eval{q'(\bar{T})}^\star$$
Thus, any abstract query $q$ that does not conforms $\mathcal{E} \lhd \eval{q(\bar{T})}^\circ$ can be safely pruned.
\end{property}

\noindent This property immediately follows \autoref{def:abstract-consistency} and \autoref{prop:relation}. Intuitively, if there exist cells in $\mathcal{E}$ that contains references from the inputs $\bar{T}$ but cannot be propagated through the abstract query $q$ with \emph{over-approximation}, then no instantiation of $q$ can enable such propagation. This property builds the foundation of the pruning algorithm.

In this way, the abstract provenance analysis returns the pruning information to the main enumerative search algorithm (\autoref{alg:top-level}) to direct the search process.

%% file: sections/experiments.tex
%!TEX root=../main.tex

\section{Experiments}
\label{sec:experiments}

To evaluate our approach, we implemented the proposed technique as a tool named \tool and tested in on 80 real world benchmarks consists of end-user analytical SQL questions and industrial database testing benchmarks. We aim to examine the following hypotheses:

\begin{itemize}
	\item \tool can solve practical analytical SQL query synthesis tasks from small demonstrations.
	\item \tool's new abstraction prunes the search space better than abstractions used in prior systems~\cite{DBLP:conf/pldi/WangCB17,DBLP:journals/pacmpl/WangFBCD20,DBLP:conf/pldi/FengMBD18}.
	%\item \tool scales better to larger input data sizes comparing against state-of-the-art techniques.
\end{itemize}

\noindent We also conducted a user study to compare users' experiences with computation demonstration and I/O examples for analytical task specification. 

\subsection{Experiment Setup}
\label{sec:experiment-setup}

\tool is implemented in Python, the user demonstration is provided via spreadsheet; however, with some additional engineering effort, \tool can work with a drag-and-drop interface~\cite{DBLP:conf/chi/WangFBDCK21} (as illustrated in \autoref{sec:overview}) for smoother experience.
%As our paper focuses on the specification and the algorithm designs, \tool follows standard program ranking and disambiguation techniques: \tool  generates queries in an order based on their complexity (queries with smaller number of subqueries first) and generates consistent programs batch-by-batch for the user to inspect until the correct query is accepted. In future, \tool can work with learning-based approaches for query ranking~\cite{DBLP:conf/icml/DevlinUBSMK17}, exploration-based~\cite{DBLP:conf/chi/WangFBDCK21} or dialog-based~\cite{DBLP:journals/pacmpl/MayerKC18,DBLP:conf/pldi/JiLXZH20} techniques for query disambiguation, which are orthogonal to our approach.
In the synthesis process, \tool enumerates join predicates (used in $\mathsf{join}$ and $\mathsf{left\_join}$) based on primary and foreign keys of the table to avoid generating unnatural predicates like ``\code{T1 Join T2 On T1.id < T2.age}''; for the same reason, \tool does not invent new constants for $\mathsf{filter}$ and considers only those provided by the user. 
%\tool supports all analytical operators in \autoref{fig:sql-grammar}. 
\tool ranks synthesized queries based on query size. %In future, it can work with more advanced ranking and ambiguity resolution frameworks~\cite{DBLP:conf/icml/DevlinUBSMK17,DBLP:conf/chi/WangFBDCK21,DBLP:journals/pacmpl/MayerKC18,DBLP:conf/pldi/JiLXZH20}.

%In our implementation, we optimize \tool to keep track of operator provenance (e.g., for deciding whether the function ``$\lambda x,y.x/y*100\%$'' can flow to an output cell $T^\circ[i,j]$) in addition to value provenance described earlier. This straightforward extension allows \tool to prune simple incorrect programs (e.g., $q_A$ in \autoref{fig:overview-enumerative-search}) with little extra overhead.

\paragraph{Benchmarks} 

We evaluate \tool on 80 benchmarks, including 60 collected from analytical SQL online tutorials/forums, and 20 tasks from TPC-DS decision support benchmark \cite{DBLP:conf/vldb/OthayothP06} --- an industry standard benchmark suite for testing commercial database systems' analytical SQL features: 
\begin{itemize}
\item From online forums and tutorials, we explored hundreds of posts and select analytical tasks with input data available. We obtained 43 easier tasks that require 2-3 operators from \autoref{fig:sql-grammar} to solve and 17 tasks that require 4-7 operators.
\item When collecting benchmarks from TPC-DS, we extract tasks by (1) isolating table view definitions (many TPC-DS tests are long sequences of tasks with one building on top of the result of another), and (2) decomposing big union queries into smaller tasks (some queries are  unions of many sub-tasks).
%, and (3) down sample the original input tables (which has thousands of rows) into smaller tables with . 
This gives us 20 benchmarks each focus on one analytical task; all requires 4-7 operators.
\end{itemize}
For each benchmark with input $\bar{T}_\mathit{raw}$ and the ground truth query $q_\mathit{gt}$, we programmatically generate small computation demonstrations using the following procedure: (1) for input with $>20$ rows, we sample 20 rows from the input table and use the sampled data $\bar{T}$ as the synthesis task input, (2) evaluate $T^\star = \eval{q_\mathit{gt}(\bar{T})}^\star$; (3) randomly sample 2 rows from $T^\star$ as $T^\star_\mathit{partial}$ and permute orders of arguments in commutative functions; (4) for expressions in $T^\star_\mathit{partial}$ with more than four values, replace it with an incomplete expression that contains at most four values together with a $\diamondsuit$ (for omitted parameters). This generates $\mathcal{E}$, a partial output table with incomplete expressions, to simulate user demonstrations that is suitable for systematic algorithm performance testing. Each benchmark is a tuple $(\bar{T}, \mathcal{E}, q_\mathit{gt})$. In total, our benchmark set contains 43 easier tasks (1-3 operators) and 37 harder tasks (4-6 operators); among them, 24 requires join , 51 tasks requires partition-aggregation and 32 requires group-aggregation. Example benchmarks can be found in the appendix.

\paragraph{Baselines}

To evaluate our algorithm, especially the provenance abstraction design, we compare \tool with two state-of-arts abstraction-based pruning techniques successfully used in relational query synthesis:
\begin{itemize}
\item \emph{Morpheus's type abstraction~\cite{DBLP:conf/pldi/FengMGDC17}:} This abstraction tracks high-level table shape information (row/column/group numbers). This technique is widely adopted in synthesizing list and table transformation programs.
\item \emph{Scythe's value abstraction~\cite{DBLP:conf/pldi/WangCB17}:} This abstraction keeps track of concrete values flow through a partial program to over-approximate its behavior. This technique is used in the state-of-the-art SQL query and visualization synthesis.
\end{itemize}
Because neither of the baselines supports analytical operators, we reimplement them by extending their abstract semantics: (1) in type abstraction rules, we extend the abstract semantics to infer the most precise table shape and group number for partition and aggregation rules, and (2) in value abstraction rules, we keep track of all known values (e.g., values from the grouping columns) for analytical operators but ignore unknown values (e.g., values from the aggregation column). Other abstraction rules for both baselines are the same as those in the original paper. We then run the evaluation using the same enumerative search framework \tool used, which keeps the search order the same for all abstractions.
%Because prior implementations of these techniques do not support analytical SQL, to better compare pruning power of these language abstractions, we extended them to support analytical SQL features and partial output data. %We reimplemented them in the same enumerative search framework used in \tool. 

\subsection{Study 1: Synthesis Efficiency}

\paragraph{Efficiency Comparison} We first test \tool's performance on practical analytical SQL tasks compared to baseline techniques. 
%We evaluate \tool on all 70 benchmarks introduced in \autoref{sec:experiment-setup} and analyze efficiency of \tool and ambiguity of the synthesized queries. 
For each benchmark $(\bar{T}, \mathcal{E}, q_\mathit{gt})$, we run \tool and two baselines with a timeout of 600 seconds. The synthesizer runs until the correct query $q_\mathit{gt}$ is found. We record (1) time each technique takes to solve the tasks, and (2) the number of consistent queries encountered by \tool (including the correct one $q_\mathit{gt}$). \tool ranks consistent queries based on the order in which their are found in the search process.  \autoref{fig:exp2-result} shows the results: each plot shows the number of benchmarks ($y$-axis) that can be solved within the given time limits ($x$-axis) by each baseline; results for easier and harder benchmarks are plotted separately.

\begin{figure}[t]
\centering
\includegraphics[width=\linewidth]{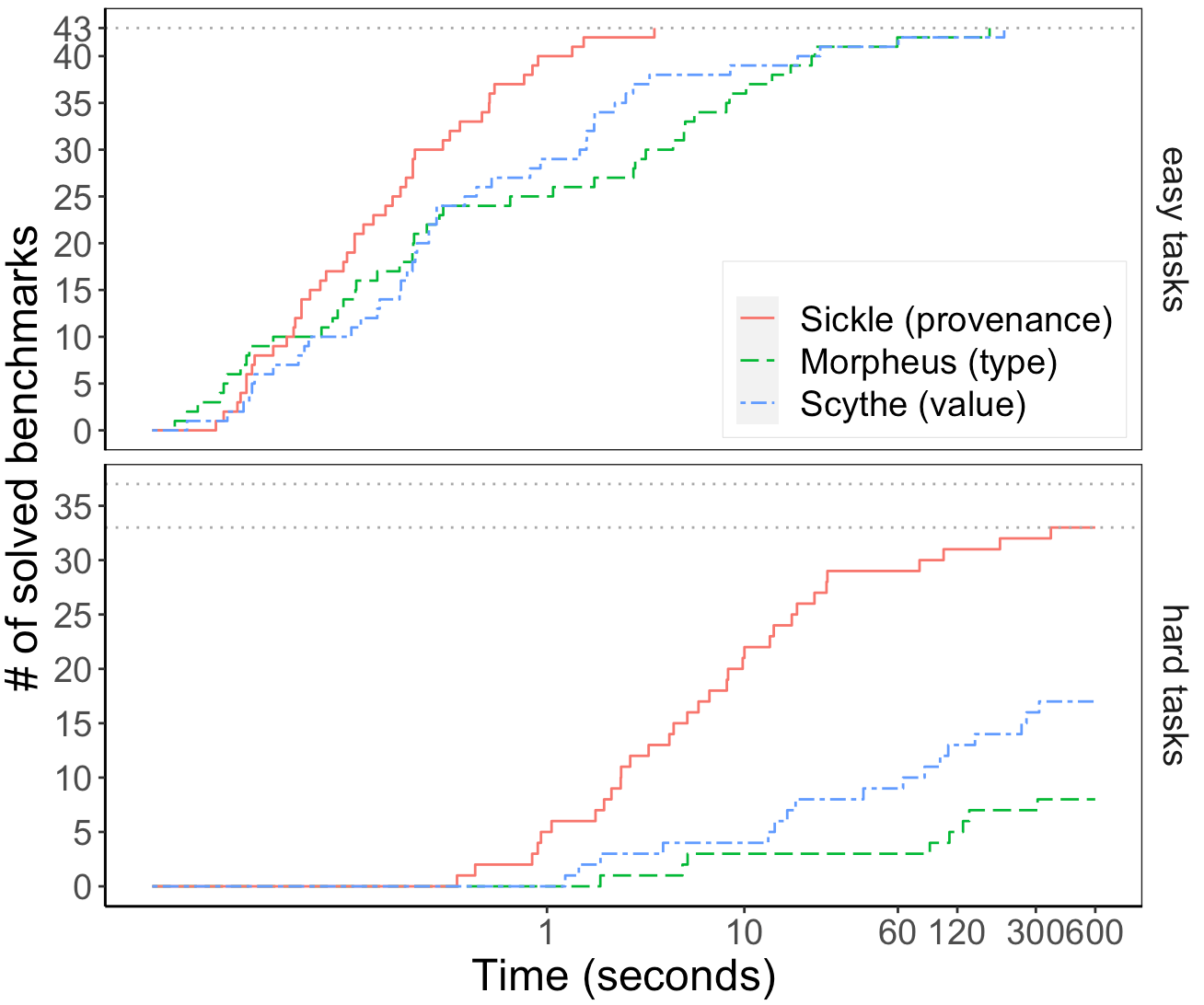}
\caption{The number of benchmarks each technique can solve within given time limit for each benchmark.}
\label{fig:exp1-result}
\end{figure}

\noindent\emph{Observation \#1: \tool can efficiently solve practical analytical SQL tasks from small demonstrations. \tool has significant advantage over baselines on harder benchmarks.}

\smallskip
As shown in \autoref{fig:exp1-result}, \tool in total solved 76 out of 80 benchmarks with a mean solving time of 12.8 seconds (\tool solved all 43 easier tasks and 33 out of 37 harder tasks). In these benchmarks, the average user demonstration size is 9 cells (the number would be 50 if full output examples are required from the user). In comparison: %\tool can solve all of the 60 benchmarks from forums and tutorials (55 of them are within 10 seconds), and the 4 unsolved cases are from the TPC-DS benchmarks. In comparison:
\begin{itemize}
    \item On the easy benchmark suite, while both baselines can also solve all 43 tasks, \tool is significantly faster: \tool's average solving time is 0.4 seconds, but type and value abstraction can take up to 8.0 and 8.6 seconds,
    \item Out of the 37 harder benchmarks, \tool solved most (33 benchmarks), and type and value abstractions solved only 8 and 17 benchmarks. \tool is also faster on tasks all techniques can solve. \tool is on average 22.3$\times$ faster than value abstraction and 69.7$\times$ faster than type abstraction.
\end{itemize}
The 4 cases \tool cannot solve are hard tasks from the TPC-DS. Upon closer inspection, cases that \tool cannot solve or are slow have the following traits: (1) the input data has many columns, or the task require \code{join} (they dramatically increase the parameter space), or (2) the solution contains many \emph{nested} operators (this makes analysis less precise). %We envision that \tool can work with learning based approaches~\cite{DBLP:conf/ijcai/EllisG17,DBLP:conf/icml/DevlinUBSMK17,DBLP:conf/iclr/KalyanMPBJG18} to prioritize search of more promising parameters to better traverse the parameter space (e.g., to decide which fields are more likely to be grouped together).

%Besides solving more benchmarks, provenance abstraction is also faster than two baseline techniques on tasks all techniques can solve: provenance abstraction is on average 46.2$\times$ faster than value abstraction and 89.0$\times$ faster than type abstraction. 

% \begin{figure}[ht]
% \begin{smpage}{0.8\linewidth}
% \centering
% \includegraphics[width=\linewidth]{images/exp1-plot.png}
% \end{smpage}

% % \bigskip

% % \begin{smpage}{\linewidth}
% % \centering
% % \begin{tabular}{cc}
% % \hline Ranking & \# of benchmarks \\\hline
% % top-1 & 62\\
% % top-3 & 62\\
% % top-5 & 63\\
% % top-10 & 65\\
% % > 10 & 66\\\hline
% % \end{tabular}
% % \end{smpage}
% \caption{The line chart that shows performance of \tool on 70 benchmarks with $x$-axis showing time limit for each benchmark (log scale), and $y$-axis showing the number of benchmarks that can be solved.
% % Right: the table that shows the number of benchmarks with the correct query ranked among top-$k$.
% }
% \label{fig:exp1-result}
% \end{figure}

\noindent\paragraph{Pruning Analysis} To understand how the provenance abstraction helps \tool to efficiently solve these practical tasks, we log and compare the number of queries (both partial and concrete queries) \tool and both baselines explored during the synthesis process. \autoref{fig:exp2-result} shows the results (top for easy tasks and bottom for hard ones). For each suite, the box plot shows the distribution of the number of queries each technique encountered when solving these benchmarks (or before timeout). We made the following observation: 

\begin{figure}[t]
\centering
\includegraphics[width=\linewidth]{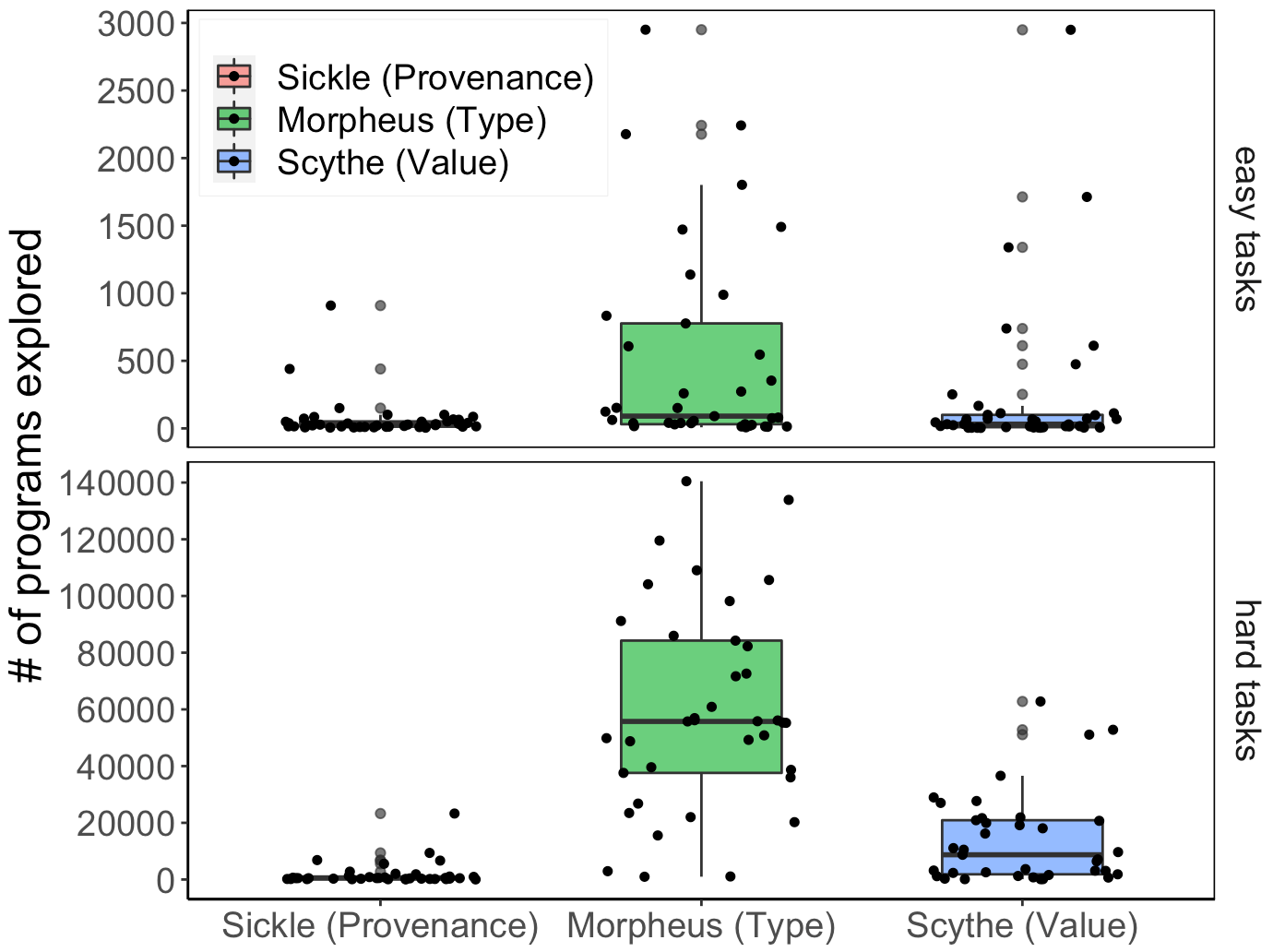}
\caption{Distributions of the number of queries explored by different techniques on easy (top) and hard (bottom) tasks.}
\label{fig:exp2-result}
\end{figure}

%We compare the pruning power of provenance abstraction with two other abstractions: type abstraction and value abstraction. We follow the same experiment process as experiment 1 and record (1) the number of tasks each technique can solve within the 600s timeout, and (2) the number of queries (both partial and concrete queries) each technique explored during the synthesis process. For better analysis, we separate results of easier benchmarks (solutions with 2-3 subqueries) from the results of harder benchmarks (query size 4-7).

\smallskip

\noindent\emph{Observation \#2: Provenance abstraction has small advantages over type and value abstractions on easier tasks because of relatively small query space to explore; but its advantage over others is significant on harder tasks with large search space.}

\smallskip

On easier tasks (\autoref{fig:exp2-result} top), despite \tool's provenance abstraction prunes the search space better, its advantage is small because these easier benchmarks require only exploring a small search space with a small number of programs inside. However, on harder tasks, the synthesizer needs to find the solution from a large search space with 100,000 --- 2,000,000 queries, and this makes provenance abstraction's pruning power shine. As shown in \autoref{fig:exp2-result}, \tool explores only 917 queries on average before finding the correct query; this is much less than type and value abstractions, which need to explore 31,371 and 6,837 queries on average (also with higher variance). 

In general, while provenance abstraction has higher reasoning overhead compared to type or value abstractions, its stronger pruning power makes it significantly outperform other techniques on challenging benchmarks as much less programs in the search space are visited. The key strength of \tool is its ability to track fine-grained provenance information through partial programs, and we envision computation demonstration and provenance abstraction to be applied to synthesis tasks in other domains rich of computation.

\smallskip

We also examined the ranking of queries synthesized by \tool to examine computation demonstration's ambiguity issues. There are 71 benchmarks with the correct query ranked at top 1, 4 cases ranked between 2 to 9, and 1 case ranked at $>$10. We noticed that computation traces and cell references are often less ambiguous than concrete value despite the smaller demonstration size; and for the case where the correct query is ranked $>10$, we found the issue is that the input data contains only one group of data --- this makes the synthesizer unable to easily generalize to multiple groups as required in the solution. This motivates that future disambiguation model could focus on seeking for more representative inputs from the user (rather than asking for more concrete/complete output demonstrations).

\smallskip

\subsection{Study 2: User Experience}

To understand users' experiences working with computation demonstration, we conducted a within-group user study with 13 participants on 6 analytical tasks. In the study, we focus on specification creation: we compare \emph{user efficiency} and \emph{user experience} using computation demonstration and output-examples for analytical task illustration. We only require the participant to understand basic computation terminologies (e.g., cumsum, rank). The 6 analysis tasks come from our benchmarks, and their complexity and input table size vary with inputs containing 10-30 rows and require 1-3 analytical operators to solve, and they are split into groups based on query complexity and table size: 3 easy tasks (1-2 analytical steps, small table), and 3 hard tasks (2-3 analytical steps, big table). Each task contains (1) a description section of the input table and the task, (2) input table, and (3) output table schema (with no information about the output content).

For each task, we assign a demonstration method that the participant needs to use: (1) output example with concrete values, (2) full computation expressions, and (3) computation demonstration with partial expressions. The participant needs to fill out three rows in the output table using the assigned specification type in a spreadsheet environment. %By focusing on specification creation in a spreadsheet environment, we are able to hide 
We counterbalance task order and specification order with the Latin square design. Each study session contains 10 minutes tutorial time, 30 minutes for solving the 6 tasks, and 20 minutes for interviews. We recorded the time each participant spent on creating the demonstration and analyzed results with Wilcoxon rank sum test; qualitative results are analyzed via inductive content analysis approach. Our study design intentionally hides the complexity of disambiguation and query understanding tasks that are orthogonal to our main contribution. This let us allocate more time to observe user behaviors on more tasks. More study details are in the supplementary material. 

\noindent\paragraph{Quantitative analysis}
In the user study, 72 out of 78 participants' solutions are correct. On the three easy tasks (1-2 analytical steps, inputs with $<15$ rows), participants are more efficient using examples (on average $122$s) compared to expressions (on average $170$s) or partial expressions (on average $168$s), but the difference is insignificant ($p>0.1$). On two of the three hard tasks, users are significantly more efficient ($p<0.05$) using expressions (on average $226$s) over examples (on average $300$s). Surprisingly, on one hard task --- ranking countries based on their average sales using the $\mathsf{Rank}$ function --- the participants are significantly more efficient using examples (on average $175$s) over expressions (on average $286$s) or partial expressions (on average $209$s) with $p < 0.05$, as we explain in our qualitative analysis.

\noindent\paragraph{Qualitative analysis} We noticed that most participants prefer expressions over examples because they enhance user confidence: participants find examples \emph{``tedious and time consuming''} (P6) and \emph{``not show how you derive the data''} (P9) but find computation demonstration make them \emph{``see what exactly I am doing''} (P1) and thus \emph{``more confident''} (P3, P7, P10). The participants also noted that they sometimes spend more time on creating expressions because they need to typeset the answers, and we envision this can be resolved with a drag-and-drop interface design in future. Second, participants mentioned that partial expressions are vital when the input is big (>20 rows). For example, \emph{``smaller expressions can be used more easily than compute the final value, and it can show the pattern''} (P5), and \emph{''I do not need to look many data or do computation on my own''} (P10). However, some participants think partial expression makes them \emph{``less confident (than full expressions), as it depends on the user's experience level to make right omission''} (P13). Finally, for the ranking task where participants perform significantly worse using expressions, participants prefer manually counting over \emph{``using an expression and go over all the data''}, because they (1) cannot easily sort the data by groups in the spreadsheet to create the full expression and (2) find omitting parameters tricky for $\mathsf{rank}$. Currently, this is a limitation of computation demonstration. We envision a mixed (example and computation demonstration) approach to make analytical tasks easy.

%% file: sections/related.tex
\section{Related Work}
\label{sec:related}

\paragraph{Relational Query Synthesis} Due to the importance of relational queries in the database industry and the programming challenges associated with them, many relational query synthesizers have been developed~\cite{DBLP:conf/sigmod/TranCP09,DBLP:conf/pldi/WangCB17,zloof1975query,DBLP:conf/pldi/ThakkarNSANR21,DBLP:conf/sigsoft/SiLZAKN18,DBLP:journals/pacmpl/RaghothamanMZNS20}. Among them, Scythe~\cite{DBLP:conf/pldi/WangCB17} supports standard SQL queries with subqueries and aggregations; EGS~\cite{DBLP:conf/pldi/ThakkarNSANR21}, ProSynth~\cite{DBLP:journals/pacmpl/RaghothamanMZNS20} and Alps~\cite{DBLP:conf/sigsoft/SiLZAKN18} are datalog query synthesizers that synthesizes datalog queries using syntax and provenance guided search from input-output examples; Morpheus~\cite{DBLP:conf/pldi/FengMGDC17} and Viser~\cite{DBLP:journals/pacmpl/WangFBCD20} are table transformation synthesizers for data wrangling.

%Because analytical SQL involves partitioning and aggregation of relational tables, logic-based induction algorithms~\cite{DBLP:conf/pldi/ThakkarNSANR21} (designed for conjunctive queries) or constraint solved-based approaches~\cite{DBLP:conf/pldi/TorlakB14,cegis2} cannot be easily adopted.

\tool's focus is the computation-rich \emph{analytical SQL}. Analytical SQL are not expressible in logic-based synthesizers~\cite{DBLP:conf/pldi/ThakkarNSANR21} that requires relational queries to be ``pure'' (free of computations); algorithms for standard SQL~\cite{DBLP:conf/pldi/WangCB17,DBLP:journals/pacmpl/WangFBCD20} are insufficient due to lack of support for analytical operators (\autoref{sec:experiments}). As \tool focuses on analytical computations, \tool is not specially designed for data wrangling~\cite{DBLP:conf/pldi/FengMGDC17} or knowledge base querying tasks~\cite{DBLP:journals/pacmpl/RaghothamanMZNS20}; there is an opportunity to combine \tool with Morpheus~\cite{sypet} or ProSynth~\cite{DBLP:journals/pacmpl/RaghothamanMZNS20} for tasks requiring all types of relational query features. \tool can also benefit from PATSQL's schema inference technique~\cite{DBLP:journals/pvldb/TakenouchiIOS21} to better scale up to larger input table by prioritizing search for columns that are type-consistent with the user demonstration.

\paragraph{Programming-by-demonstration Tools} Programming-by-example/demonstration (PBE/PBD) tools have been successfully adopted in many end user programming scenarios~\cite{DBLP:conf/popl/Gulwani11,polozov2015flashmeta,singh2012synthesizing,DBLP:conf/pldi/PolikarpovaKS16, le2014flashextract,DBLP:conf/uist/ChasinsMB18,dace,DBLP:conf/pldi/FengMGDC17,DBLP:journals/pacmpl/WangFBCD20, DBLP:journals/pacmpl/BavishiLSS21}. %PBE tools like Scythe~\cite{DBLP:conf/pldi/WangCB17} synthesize programs from input-output example pairs; 
PBD tools like Helena~\cite{DBLP:conf/uist/ChasinsMB18,DBLP:journals/pacmpl/ChasinsB17} and FlashExtract~\cite{le2014flashextract}) generalize users' data extraction actions into scripts.
%Because creating examples is more challenging in analytical tasks, we designed the new specification: computation demonstration. 
\tool's specification, computation demonstration, is also a type of PBD specification, but it differs from others: existing PBD tools synthesize scripts by ``loopifying'' the user demonstration, but our approach synthesizes a \emph{declarative query} from the demonstration. 
 For example, the user's expression sketch in DACE~\cite{DBLP:journals/pacmpl/WangDS17} can be directly used as a component of the target program, while a partial expression in \tool is only a cell-level expression generated by evaluating the target analytical query.
Thus, this domain challenge makes \tool unable to directly reuse the user demonstration as program components~\cite{DBLP:conf/pldi/CheungSM13, DBLP:journals/pacmpl/WangDS17}.
%Because computation demonstration is close to PBE specification, users who can provide I/O examples can  use it easily. 
 Gauss~\cite{DBLP:journals/pacmpl/BavishiLSS21} and \tool have similar specification design that allows the user to provide computation expressions as part of the output demonstration. Because \tool is designed to solve the analytical SQL synthesis problems where nested computation are more common (Gauss focuses on data wrangling problems), \tool makes the following different design choices to improve user experience: (1) \tool permits omitted parameters in expressions, and (2) \tool does not enforce the users' expression syntactically same as the expression appear the output from the correct query. As a result, Gauss's computation graph abstraction is too strict to apply, and \tool's choice of data provenance abstraction is more general and effectively prunes inconsistent queries during the search process.
We also envision our specification used in other computation-rich domains~\cite{DBLP:journals/tog/AhmadRCK19,DBLP:conf/pldi/SmithA16}.

Besides, many new interaction models~\cite{DBLP:conf/uist/MayerSGLMPSZG15,DBLP:conf/chi/WangFBDCK21,DBLP:conf/uist/FerdowsifardOPL20,DBLP:journals/pacmpl/PelegGIY20,DBLP:conf/pldi/JiLXZH20} are proposed make PBE and PBD tools easier-to-use. As our paper focuses on the synthesis specification and algorithm design, its interaction model is basic. As suggested by our user study, \tool could work with a drag-and-drop based interface~\cite{DBLP:conf/chi/WangFBDCK21} for easier demonstration creation dialog-based~\cite{DBLP:conf/uist/MayerSGLMPSZG15,DBLP:conf/pldi/JiLXZH20} interface for intent clarification.

%Computation demonstration is also related to PbE extensions: for example, PNbE~\cite{DBLP:conf/icse/PelegSY18,DBLP:journals/acta/PelegISY20} proposes to let users provide a program sketch to make synthesizer more efficient and better at disambiguation; SketchAX~\cite{DBLP:journals/pacmpl/AnSMS20} allows the user to provide program invariants to augment examples. Our specification differs because computation demonstration are demonstrations at the cell and expression level, not at SQL level; this lowers the specification barrier because it assumes no know analytical SQL knowledge from the user.

\paragraph{Abstraction-based Synthesis Algorithms} Abstraction-based program synthesis algorithms have been widely used in synthesis of structurally rich programs (e.g., loopy programs~\cite{DBLP:journals/pacmpl/KimHDR21}, list programs~\cite{DBLP:conf/pldi/PolikarpovaKS16,yaghmazadeh2016synthesizing}, tables programs~\cite{DBLP:conf/pldi/FengMBD18,DBLP:conf/pldi/WangCB17,DBLP:journals/pacmpl/WangFBCD20}, regex~\cite{DBLP:journals/pacmpl/WangDS18}). The core technique is to design a language abstraction~\cite{cousot1977abstract} for reasoning behaviors of partial programs and pruning ones unable to reach the goal. \tool contributes to this line of work by introducing provenance abstraction that can capture fine-grained provenance information for computation rich programs that type~\cite{DBLP:conf/pldi/FengMBD18,DBLP:conf/pldi/PolikarpovaKS16} or value-based~\cite{DBLP:conf/pldi/WangCB17,DBLP:journals/pacmpl/WangFBCD20} abstractions cannot.
While \tool currently builds on top of the enumerative search framework~\cite{Udupa:2013:TSP:2499370.2462174,DBLP:conf/pldi/FengMBD18,DBLP:conf/asplos/PhothilimthanaT16}, the abstraction can also work in solver-based ~\cite{cegis2,DBLP:conf/cav/ReynoldsBNBT19} or learning-based~\cite{DBLP:journals/pacmpl/BavishiLFSS19,DBLP:conf/icml/DevlinUBSMK17,DBLP:conf/ijcai/EllisG17} frameworks for pruning. %\tool can also work with learning-based approaches~ to further improve the search order of the programs to further improve the synthesis performance.

%The two most commonly used abstractions for table program synthesis are type abstraction~\cite{DBLP:conf/pldi/FengMGDC17,DBLP:conf/pldi/PolikarpovaKS16} and value abstraction~\cite{DBLP:conf/pldi/WangCB17,DBLP:journals/pacmpl/WangFBCD20}: the former approximates the program behavior based on high-level table shape information, and the latter leverages concrete values preserved by the partial programs. While both abstractions have been shown effective for table reshaping programs, they are less effective to capture complex (arithmetic and aggregation) computations, especially when the user output example is incomplete. Provenance abstraction used in our work leverages fine-grained provenance information to achieve more precise approximation of the program behavior that increases the pruning power.

%In SemGus~\cite{DBLP:journals/pacmpl/KimHDR21}, program abstractions are used to reason about realizability of the objective given a \emph{program prefix}; this is different from \tool, which uses program abstractions to reason about realizability of the goal given a \emph{partially filled program sketch}. This difference makes SemGus better at pruning long loopy imperative programs (with fewer parameters); while \tool are better at pruning parameter rich but short programs (suitable for analytical SQL).

%\paragraph{Interaction Models for End-user Programming} 

%% file: main.bbl

\begin{thebibliography}{47}

%%% the Web address.

\ifx \showCODEN    \undefined \def \showCODEN     #1{\unskip}     \fi
\ifx \showDOI      \undefined \def \showDOI       #1{#1}\fi
\ifx \showISBNx    \undefined \def \showISBNx     #1{\unskip}     \fi
\ifx \showISBNxiii \undefined \def \showISBNxiii  #1{\unskip}     \fi
\ifx \showISSN     \undefined \def \showISSN      #1{\unskip}     \fi
\ifx \showLCCN     \undefined \def \showLCCN      #1{\unskip}     \fi
\ifx \shownote     \undefined \def \shownote      #1{#1}          \fi
\ifx \showarticletitle \undefined \def \showarticletitle #1{#1}   \fi
\ifx \showURL      \undefined \def \showURL       {\relax}        \fi
% invisible to TeX
\providecommand\bibfield[2]{#2}
\providecommand\bibinfo[2]{#2}
\providecommand\natexlab[1]{#1}
\providecommand\showeprint[2][]{arXiv:#2}

\bibitem[\protect\citeauthoryear{Ahmad, Ragan{-}Kelley, Cheung, and
  Kamil}{Ahmad et~al\mbox{.}}{2019}]%
        {DBLP:journals/tog/AhmadRCK19}
\bibfield{author}{\bibinfo{person}{Maaz Bin~Safeer Ahmad},
  \bibinfo{person}{Jonathan Ragan{-}Kelley}, \bibinfo{person}{Alvin Cheung},
  {and} \bibinfo{person}{Shoaib Kamil}.} \bibinfo{year}{2019}\natexlab{}.
\newblock \showarticletitle{Automatically translating image processing
  libraries to halide}.
\newblock \bibinfo{journal}{\emph{{ACM} Trans. Graph.}} \bibinfo{volume}{38},
  \bibinfo{number}{6} (\bibinfo{year}{2019}), \bibinfo{pages}{204:1--204:13}.
\newblock
\urldef\tempurl%
\url{https://doi.org/10.1145/3355089.3356549}
\showDOI{\tempurl}


\bibitem[\protect\citeauthoryear{Bavishi, Lemieux, Fox, Sen, and
  Stoica}{Bavishi et~al\mbox{.}}{2019}]%
        {DBLP:journals/pacmpl/BavishiLFSS19}
\bibfield{author}{\bibinfo{person}{Rohan Bavishi}, \bibinfo{person}{Caroline
  Lemieux}, \bibinfo{person}{Roy Fox}, \bibinfo{person}{Koushik Sen}, {and}
  \bibinfo{person}{Ion Stoica}.} \bibinfo{year}{2019}\natexlab{}.
\newblock \showarticletitle{AutoPandas: neural-backed generators for program
  synthesis}.
\newblock \bibinfo{journal}{\emph{Proc. {ACM} Program. Lang.}}
  \bibinfo{volume}{3}, \bibinfo{number}{{OOPSLA}} (\bibinfo{year}{2019}),
  \bibinfo{pages}{168:1--168:27}.
\newblock
\urldef\tempurl%
\url{https://doi.org/10.1145/3360594}
\showDOI{\tempurl}


\bibitem[\protect\citeauthoryear{Bavishi, Lemieux, Sen, and Stoica}{Bavishi
  et~al\mbox{.}}{2021}]%
        {DBLP:journals/pacmpl/BavishiLSS21}
\bibfield{author}{\bibinfo{person}{Rohan Bavishi}, \bibinfo{person}{Caroline
  Lemieux}, \bibinfo{person}{Koushik Sen}, {and} \bibinfo{person}{Ion Stoica}.}
  \bibinfo{year}{2021}\natexlab{}.
\newblock \showarticletitle{Gauss: program synthesis by reasoning over graphs}.
\newblock \bibinfo{journal}{\emph{Proc. {ACM} Program. Lang.}}
  \bibinfo{volume}{5}, \bibinfo{number}{{OOPSLA}} (\bibinfo{year}{2021}),
  \bibinfo{pages}{1--29}.
\newblock
\urldef\tempurl%
\url{https://doi.org/10.1145/3485511}
\showDOI{\tempurl}


\bibitem[\protect\citeauthoryear{Chasins and Bod{\'{\i}}k}{Chasins and
  Bod{\'{\i}}k}{2017}]%
        {DBLP:journals/pacmpl/ChasinsB17}
\bibfield{author}{\bibinfo{person}{Sarah Chasins} {and}
  \bibinfo{person}{Rastislav Bod{\'{\i}}k}.} \bibinfo{year}{2017}\natexlab{}.
\newblock \showarticletitle{Skip blocks: reusing execution history to
  accelerate web scripts}.
\newblock \bibinfo{journal}{\emph{Proc. {ACM} Program. Lang.}}
  \bibinfo{volume}{1}, \bibinfo{number}{{OOPSLA}} (\bibinfo{year}{2017}),
  \bibinfo{pages}{51:1--51:28}.
\newblock
\urldef\tempurl%
\url{https://doi.org/10.1145/3133875}
\showDOI{\tempurl}


\bibitem[\protect\citeauthoryear{Chasins, Mueller, and Bod{\'{\i}}k}{Chasins
  et~al\mbox{.}}{2018}]%
        {DBLP:conf/uist/ChasinsMB18}
\bibfield{author}{\bibinfo{person}{Sarah~E. Chasins}, \bibinfo{person}{Maria
  Mueller}, {and} \bibinfo{person}{Rastislav Bod{\'{\i}}k}.}
  \bibinfo{year}{2018}\natexlab{}.
\newblock \showarticletitle{Rousillon: Scraping Distributed Hierarchical Web
  Data}. In \bibinfo{booktitle}{\emph{The 31st Annual {ACM} Symposium on User
  Interface Software and Technology, {UIST} 2018, Berlin, Germany, October
  14-17, 2018}}, \bibfield{editor}{\bibinfo{person}{Patrick Baudisch},
  \bibinfo{person}{Albrecht Schmidt}, {and} \bibinfo{person}{Andy Wilson}}
  (Eds.). \bibinfo{publisher}{{ACM}}, \bibinfo{pages}{963--975}.
\newblock
\urldef\tempurl%
\url{https://doi.org/10.1145/3242587.3242661}
\showDOI{\tempurl}


\bibitem[\protect\citeauthoryear{Chaudhuri and Dayal}{Chaudhuri and
  Dayal}{1997}]%
        {chaudhuri1997overview}
\bibfield{author}{\bibinfo{person}{Surajit Chaudhuri} {and}
  \bibinfo{person}{Umeshwar Dayal}.} \bibinfo{year}{1997}\natexlab{}.
\newblock \showarticletitle{An overview of data warehousing and OLAP
  technology}.
\newblock \bibinfo{journal}{\emph{ACM Sigmod record}} \bibinfo{volume}{26},
  \bibinfo{number}{1} (\bibinfo{year}{1997}), \bibinfo{pages}{65--74}.
\newblock


\bibitem[\protect\citeauthoryear{Cheung, Solar{-}Lezama, and Madden}{Cheung
  et~al\mbox{.}}{2013}]%
        {DBLP:conf/pldi/CheungSM13}
\bibfield{author}{\bibinfo{person}{Alvin Cheung}, \bibinfo{person}{Armando
  Solar{-}Lezama}, {and} \bibinfo{person}{Samuel Madden}.}
  \bibinfo{year}{2013}\natexlab{}.
\newblock \showarticletitle{Optimizing database-backed applications with query
  synthesis}. In \bibinfo{booktitle}{\emph{{ACM} {SIGPLAN} Conference on
  Programming Language Design and Implementation, {PLDI} '13, Seattle, WA, USA,
  June 16-19, 2013}}. \bibinfo{pages}{3--14}.
\newblock


\bibitem[\protect\citeauthoryear{Cousot and Cousot}{Cousot and Cousot}{1977}]%
        {cousot1977abstract}
\bibfield{author}{\bibinfo{person}{Patrick Cousot} {and}
  \bibinfo{person}{Radhia Cousot}.} \bibinfo{year}{1977}\natexlab{}.
\newblock \showarticletitle{Abstract interpretation: a unified lattice model
  for static analysis of programs by construction or approximation of
  fixpoints}. In \bibinfo{booktitle}{\emph{Proceedings of the 4th ACM
  SIGACT-SIGPLAN symposium on Principles of programming languages}}.
  \bibinfo{pages}{238--252}.
\newblock


\bibitem[\protect\citeauthoryear{Devlin, Uesato, Bhupatiraju, Singh, Mohamed,
  and Kohli}{Devlin et~al\mbox{.}}{2017}]%
        {DBLP:conf/icml/DevlinUBSMK17}
\bibfield{author}{\bibinfo{person}{Jacob Devlin}, \bibinfo{person}{Jonathan
  Uesato}, \bibinfo{person}{Surya Bhupatiraju}, \bibinfo{person}{Rishabh
  Singh}, \bibinfo{person}{Abdel{-}rahman Mohamed}, {and}
  \bibinfo{person}{Pushmeet Kohli}.} \bibinfo{year}{2017}\natexlab{}.
\newblock \showarticletitle{RobustFill: Neural Program Learning under Noisy
  {I/O}}. In \bibinfo{booktitle}{\emph{Proceedings of the 34th International
  Conference on Machine Learning, {ICML} 2017, Sydney, NSW, Australia, 6-11
  August 2017}}. \bibinfo{pages}{990--998}.
\newblock
\urldef\tempurl%
\url{http://proceedings.mlr.press/v70/devlin17a.html}
\showURL{%
\tempurl}


\bibitem[\protect\citeauthoryear{Ellis and Gulwani}{Ellis and Gulwani}{2017}]%
        {DBLP:conf/ijcai/EllisG17}
\bibfield{author}{\bibinfo{person}{Kevin Ellis} {and} \bibinfo{person}{Sumit
  Gulwani}.} \bibinfo{year}{2017}\natexlab{}.
\newblock \showarticletitle{Learning to Learn Programs from Examples: Going
  Beyond Program Structure}. In \bibinfo{booktitle}{\emph{Proceedings of the
  Twenty-Sixth International Joint Conference on Artificial Intelligence,
  {IJCAI} 2017, Melbourne, Australia, August 19-25, 2017}},
  \bibfield{editor}{\bibinfo{person}{Carles Sierra}} (Ed.).
  \bibinfo{publisher}{ijcai.org}, \bibinfo{pages}{1638--1645}.
\newblock
\urldef\tempurl%
\url{https://doi.org/10.24963/ijcai.2017/227}
\showDOI{\tempurl}


\bibitem[\protect\citeauthoryear{Feng, Martins, Bastani, and Dillig}{Feng
  et~al\mbox{.}}{2018}]%
        {DBLP:conf/pldi/FengMBD18}
\bibfield{author}{\bibinfo{person}{Yu Feng}, \bibinfo{person}{Ruben Martins},
  \bibinfo{person}{Osbert Bastani}, {and} \bibinfo{person}{Isil Dillig}.}
  \bibinfo{year}{2018}\natexlab{}.
\newblock \showarticletitle{Program synthesis using conflict-driven learning}.
  In \bibinfo{booktitle}{\emph{Proceedings of the 39th {ACM} {SIGPLAN}
  Conference on Programming Language Design and Implementation, {PLDI} 2018,
  Philadelphia, PA, USA, June 18-22, 2018}}. \bibinfo{pages}{420--435}.
\newblock


\bibitem[\protect\citeauthoryear{Feng, Martins, Geffen, Dillig, and
  Chaudhuri}{Feng et~al\mbox{.}}{2017a}]%
        {DBLP:conf/pldi/FengMGDC17}
\bibfield{author}{\bibinfo{person}{Yu Feng}, \bibinfo{person}{Ruben Martins},
  \bibinfo{person}{Jacob~Van Geffen}, \bibinfo{person}{Isil Dillig}, {and}
  \bibinfo{person}{Swarat Chaudhuri}.} \bibinfo{year}{2017}\natexlab{a}.
\newblock \showarticletitle{Component-based synthesis of table consolidation
  and transformation tasks from examples}. In
  \bibinfo{booktitle}{\emph{Proceedings of the 38th {ACM} {SIGPLAN} Conference
  on Programming Language Design and Implementation, {PLDI} 2017, Barcelona,
  Spain, June 18-23, 2017}}. \bibinfo{pages}{422--436}.
\newblock
\urldef\tempurl%
\url{https://doi.org/10.1145/3062341.3062351}
\showDOI{\tempurl}


\bibitem[\protect\citeauthoryear{Feng, Martins, Wang, Dillig, and Reps}{Feng
  et~al\mbox{.}}{2017b}]%
        {sypet}
\bibfield{author}{\bibinfo{person}{Yu Feng}, \bibinfo{person}{Ruben Martins},
  \bibinfo{person}{Yuepeng Wang}, \bibinfo{person}{Isil Dillig}, {and}
  \bibinfo{person}{Thomas Reps}.} \bibinfo{year}{2017}\natexlab{b}.
\newblock \showarticletitle{{Component-Based Synthesis for Complex APIs}}. In
  \bibinfo{booktitle}{\emph{Proc. Symposium on Principles of Programming
  Languages}}. ACM, \bibinfo{pages}{599--612}.
\newblock


\bibitem[\protect\citeauthoryear{Ferdowsifard, Ordookhanians, Peleg, Lerner,
  and Polikarpova}{Ferdowsifard et~al\mbox{.}}{2020}]%
        {DBLP:conf/uist/FerdowsifardOPL20}
\bibfield{author}{\bibinfo{person}{Kasra Ferdowsifard}, \bibinfo{person}{Allen
  Ordookhanians}, \bibinfo{person}{Hila Peleg}, \bibinfo{person}{Sorin Lerner},
  {and} \bibinfo{person}{Nadia Polikarpova}.} \bibinfo{year}{2020}\natexlab{}.
\newblock \showarticletitle{Small-Step Live Programming by Example}. In
  \bibinfo{booktitle}{\emph{{UIST} '20: The 33rd Annual {ACM} Symposium on User
  Interface Software and Technology, Virtual Event, USA, October 20-23, 2020}},
  \bibfield{editor}{\bibinfo{person}{Shamsi~T. Iqbal},
  \bibinfo{person}{Karon~E. MacLean}, \bibinfo{person}{Fanny Chevalier}, {and}
  \bibinfo{person}{Stefanie Mueller}} (Eds.). \bibinfo{publisher}{{ACM}},
  \bibinfo{pages}{614--626}.
\newblock
\urldef\tempurl%
\url{https://doi.org/10.1145/3379337.3415869}
\showDOI{\tempurl}


\bibitem[\protect\citeauthoryear{Gulwani}{Gulwani}{2011}]%
        {DBLP:conf/popl/Gulwani11}
\bibfield{author}{\bibinfo{person}{Sumit Gulwani}.}
  \bibinfo{year}{2011}\natexlab{}.
\newblock \showarticletitle{Automating string processing in spreadsheets using
  input-output examples}. In \bibinfo{booktitle}{\emph{Proceedings of the 38th
  {ACM} {SIGPLAN-SIGACT} Symposium on Principles of Programming Languages,
  {POPL} 2011, Austin, TX, USA, January 26-28, 2011}},
  \bibfield{editor}{\bibinfo{person}{Thomas Ball} {and} \bibinfo{person}{Mooly
  Sagiv}} (Eds.). \bibinfo{publisher}{{ACM}}, \bibinfo{pages}{317--330}.
\newblock
\urldef\tempurl%
\url{https://doi.org/10.1145/1926385.1926423}
\showDOI{\tempurl}


\bibitem[\protect\citeauthoryear{Ji, Liang, Xiong, Zhang, and Hu}{Ji
  et~al\mbox{.}}{2020}]%
        {DBLP:conf/pldi/JiLXZH20}
\bibfield{author}{\bibinfo{person}{Ruyi Ji}, \bibinfo{person}{Jingjing Liang},
  \bibinfo{person}{Yingfei Xiong}, \bibinfo{person}{Lu Zhang}, {and}
  \bibinfo{person}{Zhenjiang Hu}.} \bibinfo{year}{2020}\natexlab{}.
\newblock \showarticletitle{Question selection for interactive program
  synthesis}. In \bibinfo{booktitle}{\emph{Proceedings of the 41st {ACM}
  {SIGPLAN} International Conference on Programming Language Design and
  Implementation, {PLDI} 2020, London, UK, June 15-20, 2020}},
  \bibfield{editor}{\bibinfo{person}{Alastair~F. Donaldson} {and}
  \bibinfo{person}{Emina Torlak}} (Eds.). \bibinfo{publisher}{{ACM}},
  \bibinfo{pages}{1143--1158}.
\newblock
\urldef\tempurl%
\url{https://doi.org/10.1145/3385412.3386025}
\showDOI{\tempurl}


\bibitem[\protect\citeauthoryear{Kandel, Heer, Plaisant, Kennedy, van Ham,
  Riche, Weaver, Lee, Brodbeck, and Buono}{Kandel et~al\mbox{.}}{2011}]%
        {2011-data-wrangling}
\bibfield{author}{\bibinfo{person}{Sean Kandel}, \bibinfo{person}{Jeffrey
  Heer}, \bibinfo{person}{Catherine Plaisant}, \bibinfo{person}{Jessie
  Kennedy}, \bibinfo{person}{Frank van Ham}, \bibinfo{person}{Nathalie~Henry
  Riche}, \bibinfo{person}{Chris Weaver}, \bibinfo{person}{Bongshin Lee},
  \bibinfo{person}{Dominique Brodbeck}, {and} \bibinfo{person}{Paolo Buono}.}
  \bibinfo{year}{2011}\natexlab{}.
\newblock \showarticletitle{Research Directions in Data Wrangling:
  Visualizations and Transformations for Usable and Credible Data}.
\newblock \bibinfo{journal}{\emph{Information Visualization Journal}}
  \bibinfo{volume}{10}, \bibinfo{number}{4} (\bibinfo{year}{2011}),
  \bibinfo{pages}{271--288}.
\newblock


\bibitem[\protect\citeauthoryear{Kandel, Paepcke, Hellerstein, and Heer}{Kandel
  et~al\mbox{.}}{2012}]%
        {2012-enterprise-analysis-interviews}
\bibfield{author}{\bibinfo{person}{Sean Kandel}, \bibinfo{person}{Andreas
  Paepcke}, \bibinfo{person}{Joseph Hellerstein}, {and}
  \bibinfo{person}{Jeffrey Heer}.} \bibinfo{year}{2012}\natexlab{}.
\newblock \showarticletitle{Enterprise Data Analysis and Visualization: An
  Interview Study}. In \bibinfo{booktitle}{\emph{IEEE Visual Analytics Science
  \& Technology (VAST)}}.
\newblock
\urldef\tempurl%
\url{http://idl.cs.washington.edu/papers/enterprise-analysis-interviews}
\showURL{%
\tempurl}


\bibitem[\protect\citeauthoryear{Kim, Hu, D'Antoni, and Reps}{Kim
  et~al\mbox{.}}{2021}]%
        {DBLP:journals/pacmpl/KimHDR21}
\bibfield{author}{\bibinfo{person}{Jinwoo Kim}, \bibinfo{person}{Qinheping Hu},
  \bibinfo{person}{Loris D'Antoni}, {and} \bibinfo{person}{Thomas~W. Reps}.}
  \bibinfo{year}{2021}\natexlab{}.
\newblock \showarticletitle{Semantics-guided synthesis}.
\newblock \bibinfo{journal}{\emph{Proc. {ACM} Program. Lang.}}
  \bibinfo{volume}{5}, \bibinfo{number}{{POPL}} (\bibinfo{year}{2021}),
  \bibinfo{pages}{1--32}.
\newblock
\urldef\tempurl%
\url{https://doi.org/10.1145/3434311}
\showDOI{\tempurl}


\bibitem[\protect\citeauthoryear{Le and Gulwani}{Le and Gulwani}{2014}]%
        {le2014flashextract}
\bibfield{author}{\bibinfo{person}{Vu Le} {and} \bibinfo{person}{Sumit
  Gulwani}.} \bibinfo{year}{2014}\natexlab{}.
\newblock \showarticletitle{FlashExtract: A framework for data extraction by
  examples}. In \bibinfo{booktitle}{\emph{ACM SIGPLAN Notices}},
  Vol.~\bibinfo{volume}{49}. ACM.
\newblock


\bibitem[\protect\citeauthoryear{Mayer, Soares, Grechkin, Le, Marron, Polozov,
  Singh, Zorn, and Gulwani}{Mayer et~al\mbox{.}}{2015}]%
        {DBLP:conf/uist/MayerSGLMPSZG15}
\bibfield{author}{\bibinfo{person}{Mika{\"{e}}l Mayer},
  \bibinfo{person}{Gustavo Soares}, \bibinfo{person}{Maxim Grechkin},
  \bibinfo{person}{Vu Le}, \bibinfo{person}{Mark Marron},
  \bibinfo{person}{Oleksandr Polozov}, \bibinfo{person}{Rishabh Singh},
  \bibinfo{person}{Benjamin~G. Zorn}, {and} \bibinfo{person}{Sumit Gulwani}.}
  \bibinfo{year}{2015}\natexlab{}.
\newblock \showarticletitle{User Interaction Models for Disambiguation in
  Programming by Example}. In \bibinfo{booktitle}{\emph{Proceedings of the 28th
  Annual {ACM} Symposium on User Interface Software {\&} Technology, {UIST}
  2015, Charlotte, NC, USA, November 8-11, 2015}},
  \bibfield{editor}{\bibinfo{person}{Celine Latulipe}, \bibinfo{person}{Bjoern
  Hartmann}, {and} \bibinfo{person}{Tovi Grossman}} (Eds.).
  \bibinfo{publisher}{{ACM}}, \bibinfo{pages}{291--301}.
\newblock
\urldef\tempurl%
\url{https://doi.org/10.1145/2807442.2807459}
\showDOI{\tempurl}


\bibitem[\protect\citeauthoryear{Miltner, Fisher, Pierce, Walker, and
  Zdancewic}{Miltner et~al\mbox{.}}{2018}]%
        {DBLP:journals/pacmpl/MiltnerFPWZ18}
\bibfield{author}{\bibinfo{person}{Anders Miltner}, \bibinfo{person}{Kathleen
  Fisher}, \bibinfo{person}{Benjamin~C. Pierce}, \bibinfo{person}{David
  Walker}, {and} \bibinfo{person}{Steve Zdancewic}.}
  \bibinfo{year}{2018}\natexlab{}.
\newblock \showarticletitle{Synthesizing bijective lenses}.
\newblock \bibinfo{journal}{\emph{{PACMPL}}} \bibinfo{volume}{2},
  \bibinfo{number}{{POPL}} (\bibinfo{year}{2018}), \bibinfo{pages}{1:1--1:30}.
\newblock
\urldef\tempurl%
\url{https://doi.org/10.1145/3158089}
\showDOI{\tempurl}


\bibitem[\protect\citeauthoryear{Nambiar and Poess}{Nambiar and Poess}{2006}]%
        {DBLP:conf/vldb/OthayothP06}
\bibfield{author}{\bibinfo{person}{Raghunath~Othayoth Nambiar} {and}
  \bibinfo{person}{Meikel Poess}.} \bibinfo{year}{2006}\natexlab{}.
\newblock \showarticletitle{The Making of {TPC-DS}}. In
  \bibinfo{booktitle}{\emph{Proceedings of the 32nd International Conference on
  Very Large Data Bases, Seoul, Korea, September 12-15, 2006}},
  \bibfield{editor}{\bibinfo{person}{Umeshwar Dayal},
  \bibinfo{person}{Kyu{-}Young Whang}, \bibinfo{person}{David~B. Lomet},
  \bibinfo{person}{Gustavo Alonso}, \bibinfo{person}{Guy~M. Lohman},
  \bibinfo{person}{Martin~L. Kersten}, \bibinfo{person}{Sang~Kyun Cha}, {and}
  \bibinfo{person}{Young{-}Kuk Kim}} (Eds.). \bibinfo{publisher}{{ACM}},
  \bibinfo{pages}{1049--1058}.
\newblock
\urldef\tempurl%
\url{http://dl.acm.org/citation.cfm?id=1164217}
\showURL{%
\tempurl}


\bibitem[\protect\citeauthoryear{Parameswaran}{Parameswaran}{2019}]%
        {parameswaran2019enabling}
\bibfield{author}{\bibinfo{person}{Aditya Parameswaran}.}
  \bibinfo{year}{2019}\natexlab{}.
\newblock \showarticletitle{Enabling data science for the majority}.
\newblock \bibinfo{journal}{\emph{Proceedings of the VLDB Endowment}}
  \bibinfo{volume}{12}, \bibinfo{number}{12} (\bibinfo{year}{2019}),
  \bibinfo{pages}{2309--2322}.
\newblock


\bibitem[\protect\citeauthoryear{Peleg, Gabay, Itzhaky, and Yahav}{Peleg
  et~al\mbox{.}}{2020}]%
        {DBLP:journals/pacmpl/PelegGIY20}
\bibfield{author}{\bibinfo{person}{Hila Peleg}, \bibinfo{person}{Roi Gabay},
  \bibinfo{person}{Shachar Itzhaky}, {and} \bibinfo{person}{Eran Yahav}.}
  \bibinfo{year}{2020}\natexlab{}.
\newblock \showarticletitle{Programming with a read-eval-synth loop}.
\newblock \bibinfo{journal}{\emph{Proc. {ACM} Program. Lang.}}
  \bibinfo{volume}{4}, \bibinfo{number}{{OOPSLA}} (\bibinfo{year}{2020}),
  \bibinfo{pages}{159:1--159:30}.
\newblock
\urldef\tempurl%
\url{https://doi.org/10.1145/3428227}
\showDOI{\tempurl}


\bibitem[\protect\citeauthoryear{Phothilimthana, Thakur, Bod{\'{\i}}k, and
  Dhurjati}{Phothilimthana et~al\mbox{.}}{2016}]%
        {DBLP:conf/asplos/PhothilimthanaT16}
\bibfield{author}{\bibinfo{person}{Phitchaya~Mangpo Phothilimthana},
  \bibinfo{person}{Aditya Thakur}, \bibinfo{person}{Rastislav Bod{\'{\i}}k},
  {and} \bibinfo{person}{Dinakar Dhurjati}.} \bibinfo{year}{2016}\natexlab{}.
\newblock \showarticletitle{Scaling up Superoptimization}. In
  \bibinfo{booktitle}{\emph{{ASPLOS} '16, Atlanta, GA, USA, April 2-6, 2016}}.
  \bibinfo{pages}{297--310}.
\newblock


\bibitem[\protect\citeauthoryear{Polikarpova, Kuraj, and
  Solar{-}Lezama}{Polikarpova et~al\mbox{.}}{2016}]%
        {DBLP:conf/pldi/PolikarpovaKS16}
\bibfield{author}{\bibinfo{person}{Nadia Polikarpova}, \bibinfo{person}{Ivan
  Kuraj}, {and} \bibinfo{person}{Armando Solar{-}Lezama}.}
  \bibinfo{year}{2016}\natexlab{}.
\newblock \showarticletitle{Program synthesis from polymorphic refinement
  types}. In \bibinfo{booktitle}{\emph{Proceedings of the 37th {ACM} {SIGPLAN}
  Conference on Programming Language Design and Implementation, {PLDI} 2016,
  Santa Barbara, CA, USA, June 13-17, 2016}}. \bibinfo{pages}{522--538}.
\newblock
\urldef\tempurl%
\url{https://doi.org/10.1145/2908080.2908093}
\showDOI{\tempurl}


\bibitem[\protect\citeauthoryear{Polozov and Gulwani}{Polozov and
  Gulwani}{2015}]%
        {polozov2015flashmeta}
\bibfield{author}{\bibinfo{person}{Oleksandr Polozov} {and}
  \bibinfo{person}{Sumit Gulwani}.} \bibinfo{year}{2015}\natexlab{}.
\newblock \showarticletitle{FlashMeta: A framework for inductive program
  synthesis}. In \bibinfo{booktitle}{\emph{Proceedings of the 2015 ACM SIGPLAN
  International Conference on Object-Oriented Programming, Systems, Languages,
  and Applications}}. ACM, \bibinfo{pages}{107--126}.
\newblock


\bibitem[\protect\citeauthoryear{Raghothaman, Mendelson, Zhao, Naik, and
  Scholz}{Raghothaman et~al\mbox{.}}{2020}]%
        {DBLP:journals/pacmpl/RaghothamanMZNS20}
\bibfield{author}{\bibinfo{person}{Mukund Raghothaman},
  \bibinfo{person}{Jonathan Mendelson}, \bibinfo{person}{David Zhao},
  \bibinfo{person}{Mayur Naik}, {and} \bibinfo{person}{Bernhard Scholz}.}
  \bibinfo{year}{2020}\natexlab{}.
\newblock \showarticletitle{Provenance-guided synthesis of Datalog programs}.
\newblock \bibinfo{journal}{\emph{Proc. {ACM} Program. Lang.}}
  \bibinfo{volume}{4}, \bibinfo{number}{{POPL}} (\bibinfo{year}{2020}),
  \bibinfo{pages}{62:1--62:27}.
\newblock
\urldef\tempurl%
\url{https://doi.org/10.1145/3371130}
\showDOI{\tempurl}


\bibitem[\protect\citeauthoryear{Reynolds, Barbosa, N{\"{o}}tzli, Barrett, and
  Tinelli}{Reynolds et~al\mbox{.}}{2019}]%
        {DBLP:conf/cav/ReynoldsBNBT19}
\bibfield{author}{\bibinfo{person}{Andrew Reynolds}, \bibinfo{person}{Haniel
  Barbosa}, \bibinfo{person}{Andres N{\"{o}}tzli}, \bibinfo{person}{Clark~W.
  Barrett}, {and} \bibinfo{person}{Cesare Tinelli}.}
  \bibinfo{year}{2019}\natexlab{}.
\newblock \showarticletitle{cvc4sy: Smart and Fast Term Enumeration for
  Syntax-Guided Synthesis}. In \bibinfo{booktitle}{\emph{Computer Aided
  Verification - 31st International Conference, {CAV} 2019, New York City, NY,
  USA, July 15-18, 2019,}}, \bibfield{editor}{\bibinfo{person}{Isil Dillig}
  {and} \bibinfo{person}{Serdar Tasiran}} (Eds.), Vol.~\bibinfo{volume}{11562}.
  \bibinfo{publisher}{Springer}, \bibinfo{pages}{74--83}.
\newblock
\urldef\tempurl%
\url{https://doi.org/10.1007/978-3-030-25543-5\_5}
\showURL{%
\tempurl}


\bibitem[\protect\citeauthoryear{Si, Lee, Zhang, Albarghouthi, Koutris, and
  Naik}{Si et~al\mbox{.}}{2018}]%
        {DBLP:conf/sigsoft/SiLZAKN18}
\bibfield{author}{\bibinfo{person}{Xujie Si}, \bibinfo{person}{Woosuk Lee},
  \bibinfo{person}{Richard Zhang}, \bibinfo{person}{Aws Albarghouthi},
  \bibinfo{person}{Paraschos Koutris}, {and} \bibinfo{person}{Mayur Naik}.}
  \bibinfo{year}{2018}\natexlab{}.
\newblock \showarticletitle{Syntax-guided synthesis of Datalog programs}. In
  \bibinfo{booktitle}{\emph{{ESEC/SIGSOFT} {FSE} 2018}}.
  \bibinfo{pages}{515--527}.
\newblock


\bibitem[\protect\citeauthoryear{Singh and Gulwani}{Singh and Gulwani}{2012}]%
        {singh2012synthesizing}
\bibfield{author}{\bibinfo{person}{Rishabh Singh} {and} \bibinfo{person}{Sumit
  Gulwani}.} \bibinfo{year}{2012}\natexlab{}.
\newblock \showarticletitle{Synthesizing number transformations from
  input-output examples}. In \bibinfo{booktitle}{\emph{International Conference
  on Computer Aided Verification}}. Springer, \bibinfo{pages}{634--651}.
\newblock


\bibitem[\protect\citeauthoryear{Smith and Albarghouthi}{Smith and
  Albarghouthi}{2016}]%
        {DBLP:conf/pldi/SmithA16}
\bibfield{author}{\bibinfo{person}{Calvin Smith} {and} \bibinfo{person}{Aws
  Albarghouthi}.} \bibinfo{year}{2016}\natexlab{}.
\newblock \showarticletitle{MapReduce program synthesis}. In
  \bibinfo{booktitle}{\emph{Proceedings of the 37th {ACM} {SIGPLAN} Conference
  on Programming Language Design and Implementation, {PLDI} 2016, Santa
  Barbara, CA, USA, June 13-17, 2016}},
  \bibfield{editor}{\bibinfo{person}{Chandra Krintz} {and}
  \bibinfo{person}{Emery Berger}} (Eds.). \bibinfo{publisher}{{ACM}},
  \bibinfo{pages}{326--340}.
\newblock
\urldef\tempurl%
\url{https://doi.org/10.1145/2908080.2908102}
\showDOI{\tempurl}


\bibitem[\protect\citeauthoryear{Solar-Lezama}{Solar-Lezama}{2009}]%
        {cegis2}
\bibfield{author}{\bibinfo{person}{Armando Solar-Lezama}.}
  \bibinfo{year}{2009}\natexlab{}.
\newblock \showarticletitle{The Sketching Approach to Program Synthesis.}. In
  \bibinfo{booktitle}{\emph{Proc. Asian Symposium on Programming Languages and
  Systems}}. Springer, \bibinfo{pages}{4--13}.
\newblock


\bibitem[\protect\citeauthoryear{Takenouchi, Ishio, Okada, and
  Sakata}{Takenouchi et~al\mbox{.}}{2021}]%
        {DBLP:journals/pvldb/TakenouchiIOS21}
\bibfield{author}{\bibinfo{person}{Keita Takenouchi}, \bibinfo{person}{Takashi
  Ishio}, \bibinfo{person}{Joji Okada}, {and} \bibinfo{person}{Yuji Sakata}.}
  \bibinfo{year}{2021}\natexlab{}.
\newblock \showarticletitle{{PATSQL:} Efficient Synthesis of {SQL} Queries from
  Example Tables with Quick Inference of Projected Columns}.
\newblock \bibinfo{journal}{\emph{Proc. {VLDB} Endow.}} \bibinfo{volume}{14},
  \bibinfo{number}{11} (\bibinfo{year}{2021}), \bibinfo{pages}{1937--1949}.
\newblock
\urldef\tempurl%
\url{http://www.vldb.org/pvldb/vol14/p1937-takenouchi.pdf}
\showURL{%
\tempurl}


\bibitem[\protect\citeauthoryear{Thakkar, Naik, Sands, Alur, Naik, and
  Raghothaman}{Thakkar et~al\mbox{.}}{2021}]%
        {DBLP:conf/pldi/ThakkarNSANR21}
\bibfield{author}{\bibinfo{person}{Aalok Thakkar}, \bibinfo{person}{Aaditya
  Naik}, \bibinfo{person}{Nathaniel Sands}, \bibinfo{person}{Rajeev Alur},
  \bibinfo{person}{Mayur Naik}, {and} \bibinfo{person}{Mukund Raghothaman}.}
  \bibinfo{year}{2021}\natexlab{}.
\newblock \showarticletitle{Example-guided synthesis of relational queries}. In
  \bibinfo{booktitle}{\emph{{PLDI} '21, Virtual Event, Canada, June 20-25,
  20211}}, \bibfield{editor}{\bibinfo{person}{Stephen~N. Freund} {and}
  \bibinfo{person}{Eran Yahav}} (Eds.). \bibinfo{publisher}{{ACM}},
  \bibinfo{pages}{1110--1125}.
\newblock


\bibitem[\protect\citeauthoryear{Tran, Chan, and Parthasarathy}{Tran
  et~al\mbox{.}}{2009}]%
        {DBLP:conf/sigmod/TranCP09}
\bibfield{author}{\bibinfo{person}{Quoc~Trung Tran},
  \bibinfo{person}{Chee{-}Yong Chan}, {and} \bibinfo{person}{Srinivasan
  Parthasarathy}.} \bibinfo{year}{2009}\natexlab{}.
\newblock \showarticletitle{Query by output}. In
  \bibinfo{booktitle}{\emph{{SIGMOD} 2009, Providence, Rhode Island, USA, June
  29 - July 2, 2009}}. \bibinfo{pages}{535--548}.
\newblock
\urldef\tempurl%
\url{https://doi.org/10.1145/1559845.1559902}
\showDOI{\tempurl}


\bibitem[\protect\citeauthoryear{Udupa, Raghavan, Deshmukh, Mador-Haim, Martin,
  and Alur}{Udupa et~al\mbox{.}}{2013}]%
        {Udupa:2013:TSP:2499370.2462174}
\bibfield{author}{\bibinfo{person}{Abhishek Udupa}, \bibinfo{person}{Arun
  Raghavan}, \bibinfo{person}{Jyotirmoy~V. Deshmukh}, \bibinfo{person}{Sela
  Mador-Haim}, \bibinfo{person}{Milo~M.K. Martin}, {and}
  \bibinfo{person}{Rajeev Alur}.} \bibinfo{year}{2013}\natexlab{}.
\newblock \showarticletitle{TRANSIT: Specifying Protocols with Concolic
  Snippets}.
\newblock \bibinfo{journal}{\emph{SIGPLAN Not.}} \bibinfo{volume}{48},
  \bibinfo{number}{6} (\bibinfo{date}{June} \bibinfo{year}{2013}),
  \bibinfo{pages}{287--296}.
\newblock
\showISSN{0362-1340}
\urldef\tempurl%
\url{https://doi.org/10.1145/2499370.2462174}
\showDOI{\tempurl}


\bibitem[\protect\citeauthoryear{Wang, Cheung, and Bod{\'{\i}}k}{Wang
  et~al\mbox{.}}{2017a}]%
        {DBLP:conf/pldi/WangCB17}
\bibfield{author}{\bibinfo{person}{Chenglong Wang}, \bibinfo{person}{Alvin
  Cheung}, {and} \bibinfo{person}{Rastislav Bod{\'{\i}}k}.}
  \bibinfo{year}{2017}\natexlab{a}.
\newblock \showarticletitle{Synthesizing highly expressive {SQL} queries from
  input-output examples}. In \bibinfo{booktitle}{\emph{Proceedings of the 38th
  {ACM} {SIGPLAN} Conference on Programming Language Design and Implementation,
  {PLDI} 2017, Barcelona, Spain, June 18-23, 2017}}. \bibinfo{pages}{452--466}.
\newblock
\urldef\tempurl%
\url{https://doi.org/10.1145/3062341.3062365}
\showDOI{\tempurl}


\bibitem[\protect\citeauthoryear{Wang, Feng, Bod{\'{\i}}k, Cheung, and
  Dillig}{Wang et~al\mbox{.}}{2020}]%
        {DBLP:journals/pacmpl/WangFBCD20}
\bibfield{author}{\bibinfo{person}{Chenglong Wang}, \bibinfo{person}{Yu Feng},
  \bibinfo{person}{Rastislav Bod{\'{\i}}k}, \bibinfo{person}{Alvin Cheung},
  {and} \bibinfo{person}{Isil Dillig}.} \bibinfo{year}{2020}\natexlab{}.
\newblock \showarticletitle{Visualization by example}.
\newblock \bibinfo{journal}{\emph{Proc. {ACM} Program. Lang.}}
  \bibinfo{volume}{4}, \bibinfo{number}{{POPL}} (\bibinfo{year}{2020}),
  \bibinfo{pages}{49:1--49:28}.
\newblock
\urldef\tempurl%
\url{https://doi.org/10.1145/3371117}
\showDOI{\tempurl}


\bibitem[\protect\citeauthoryear{Wang, Feng, Bod{\'{\i}}k, Dillig, Cheung, and
  Ko}{Wang et~al\mbox{.}}{2021}]%
        {DBLP:conf/chi/WangFBDCK21}
\bibfield{author}{\bibinfo{person}{Chenglong Wang}, \bibinfo{person}{Yu Feng},
  \bibinfo{person}{Rastislav Bod{\'{\i}}k}, \bibinfo{person}{Isil Dillig},
  \bibinfo{person}{Alvin Cheung}, {and} \bibinfo{person}{Amy~J. Ko}.}
  \bibinfo{year}{2021}\natexlab{}.
\newblock \showarticletitle{Falx: Synthesis-Powered Visualization Authoring}.
  In \bibinfo{booktitle}{\emph{{CHI} '21, Virtual Event / Yokohama, Japan, May
  8-13, 2021}}, \bibfield{editor}{\bibinfo{person}{Yoshifumi Kitamura},
  \bibinfo{person}{Aaron Quigley}, \bibinfo{person}{Katherine Isbister},
  \bibinfo{person}{Takeo Igarashi}, \bibinfo{person}{Pernille Bj{\o}rn}, {and}
  \bibinfo{person}{Steven~Mark Drucker}} (Eds.). \bibinfo{publisher}{{ACM}},
  \bibinfo{pages}{106:1--106:15}.
\newblock
\urldef\tempurl%
\url{https://doi.org/10.1145/3411764.3445249}
\showDOI{\tempurl}


\bibitem[\protect\citeauthoryear{Wang, Dillig, and Singh}{Wang
  et~al\mbox{.}}{2017b}]%
        {dace}
\bibfield{author}{\bibinfo{person}{Xinyu Wang}, \bibinfo{person}{Isil Dillig},
  {and} \bibinfo{person}{Rishabh Singh}.} \bibinfo{year}{2017}\natexlab{b}.
\newblock \showarticletitle{{Synthesis of Data Completion Scripts using Finite
  Tree Automata}}. In \bibinfo{booktitle}{\emph{Proc. International Conference
  on Object-Oriented Programming, Systems, Languages, and Applications}}.
  \bibinfo{publisher}{ACM}, \bibinfo{pages}{62:1--62:26}.
\newblock


\bibitem[\protect\citeauthoryear{Wang, Dillig, and Singh}{Wang
  et~al\mbox{.}}{2017c}]%
        {DBLP:journals/pacmpl/WangDS17}
\bibfield{author}{\bibinfo{person}{Xinyu Wang}, \bibinfo{person}{Isil Dillig},
  {and} \bibinfo{person}{Rishabh Singh}.} \bibinfo{year}{2017}\natexlab{c}.
\newblock \showarticletitle{Synthesis of data completion scripts using finite
  tree automata}.
\newblock \bibinfo{journal}{\emph{{PACMPL}}} \bibinfo{volume}{1},
  \bibinfo{number}{{OOPSLA}} (\bibinfo{year}{2017}),
  \bibinfo{pages}{62:1--62:26}.
\newblock
\urldef\tempurl%
\url{https://doi.org/10.1145/3133886}
\showDOI{\tempurl}


\bibitem[\protect\citeauthoryear{Wang, Dillig, and Singh}{Wang
  et~al\mbox{.}}{2018}]%
        {DBLP:journals/pacmpl/WangDS18}
\bibfield{author}{\bibinfo{person}{Xinyu Wang}, \bibinfo{person}{Isil Dillig},
  {and} \bibinfo{person}{Rishabh Singh}.} \bibinfo{year}{2018}\natexlab{}.
\newblock \showarticletitle{Program synthesis using abstraction refinement}.
\newblock \bibinfo{journal}{\emph{Proc. {ACM} Program. Lang.}}
  \bibinfo{volume}{2}, \bibinfo{number}{{POPL}} (\bibinfo{year}{2018}),
  \bibinfo{pages}{63:1--63:30}.
\newblock
\urldef\tempurl%
\url{https://doi.org/10.1145/3158151}
\showDOI{\tempurl}


\bibitem[\protect\citeauthoryear{Willsey, Nandi, Wang, Flatt, Tatlock, and
  Panchekha}{Willsey et~al\mbox{.}}{2021}]%
        {willsey2021egg}
\bibfield{author}{\bibinfo{person}{Max Willsey}, \bibinfo{person}{Chandrakana
  Nandi}, \bibinfo{person}{Yisu~Remy Wang}, \bibinfo{person}{Oliver Flatt},
  \bibinfo{person}{Zachary Tatlock}, {and} \bibinfo{person}{Pavel Panchekha}.}
  \bibinfo{year}{2021}\natexlab{}.
\newblock \showarticletitle{Egg: Fast and extensible equality saturation}.
\newblock \bibinfo{journal}{\emph{Proceedings of the ACM on Programming
  Languages}} \bibinfo{volume}{5}, \bibinfo{number}{POPL}
  (\bibinfo{year}{2021}), \bibinfo{pages}{1--29}.
\newblock


\bibitem[\protect\citeauthoryear{Yaghmazadeh, Klinger, Dillig, and
  Chaudhuri}{Yaghmazadeh et~al\mbox{.}}{2016}]%
        {yaghmazadeh2016synthesizing}
\bibfield{author}{\bibinfo{person}{Navid Yaghmazadeh},
  \bibinfo{person}{Christian Klinger}, \bibinfo{person}{Isil Dillig}, {and}
  \bibinfo{person}{Swarat Chaudhuri}.} \bibinfo{year}{2016}\natexlab{}.
\newblock \showarticletitle{Synthesizing transformations on hierarchically
  structured data}. In \bibinfo{booktitle}{\emph{Proceedings of the 37th ACM
  SIGPLAN Conference on Programming Language Design and Implementation}}. ACM,
  \bibinfo{pages}{508--521}.
\newblock


\bibitem[\protect\citeauthoryear{Zloof}{Zloof}{1975}]%
        {zloof1975query}
\bibfield{author}{\bibinfo{person}{Mosh{\'e}~M Zloof}.}
  \bibinfo{year}{1975}\natexlab{}.
\newblock \showarticletitle{Query by example}. In
  \bibinfo{booktitle}{\emph{Proceedings of the May 19-22, 1975, national
  computer conference and exposition}}. ACM, \bibinfo{pages}{431--438}.
\newblock


\end{thebibliography}
